\documentclass[letterpaper]{JHEP3}

\usepackage{amssymb}
\usepackage{amsmath}
\usepackage{bbold}
\usepackage{graphicx}

\newcommand{\bea}{\begin{eqnarray}}
\newcommand{\eea}{\end{eqnarray}}
\newcommand{\beq}{\begin{equation}}
\newcommand{\eeq}{\end{equation}}
\newcommand{\nn}{\nonumber}
\newcommand{\ignore}[1]{}

\def\<{\langle}
\def\>{\rangle}

\def\Hp{H_{\rm p}}
\def\Hd{H_{\rm d}}
\def\HL{H_\Lambda}

\def\r{\overline{r}}
\def\t{\overline{t}}
\def\a{\overline{a}}

\def\tobs{\tau_{\rm obs}}
\def\tc{\tau_{\rm c}}
\def\Nc{N_\Lambda}

\title{Phenomenology of the CAH+ measure}

\author{Michael P.~Salem$^{a}$ and Alexander Vilenkin$^{b}$\\
$^a$Department of Physics, Stanford University, Stanford, 
California 94305, USA\\
$^b$Institute of Cosmology, Department of Physics and Astronomy,  
Tufts University,\\ \,\,\,Medford, Massachusetts 02155, USA\\
}

\abstract{The CAH+ measure regulates the infinite spacetime volume of 
the multiverse by constructing a surface of constant comoving apparent 
horizon (CAH) and then removing the future lightcones of all points on 
that surface (the latter prescription is referred to by the ``+'' in the 
name of the measure).  This measure was motivated by the conjectured 
duality between the bulk of the multiverse and its future infinity and 
by the causality condition, requiring that the cutoff surfaces of the 
measure should be spacelike or null.  Here we investigate the 
phenomenology of the CAH+ measure and find that it does not suffer from 
any known pathologies.  The distribution for the cosmological constant 
$\Lambda$ derived from this measure is in a good agreement with the 
observed value, and the distribution for the number of inflationary 
$e$-foldings satisfies the observational constraint.  The CAH+ measure 
does not exhibit any ``runaway'' behaviors at zero or negative values 
of Lambda, which have been recently shown to afflict a number of other 
measures.}

\preprint{SU-ITP-11/35}

\begin{document}

\section{Introduction} 
\label{sec:introduction}

The most persistent unresolved problem of inflationary 
cosmology is the measure problem.  The crux of the problem is that 
the numbers of all kinds of events occurring over the course 
of eternal inflation grow exponentially with time and become 
infinite in the limit.  Whatever cutoff method is used to regulate 
these infinities, most of the events occur close to the cutoff, so 
the resulting probability measure depends sensitively on the cutoff 
prescription.

A number of different measures have been proposed, and many of
their properties have been investigated.  (For recent discussion,
including references, see for example \cite{DeSimone:2008if,BFLR,
Linde:2008xf,Nomura,Freivogel:2011eg}.)  This work has shown that 
some of the proposals are not viable, since they lead to paradoxes 
or to conflict with observations.  It seems unlikely, however, that 
this kind of phenomenological analysis will yield a unique 
prescription for the measure. 

The choice of measure should ultimately be determined by the underlying 
fundamental theory.  In this spirit, it was proposed in 
\cite{holographic1,holographic2} that the dynamics of the inflationary 
multiverse has a dual description in the form of a lower-dimensional 
Euclidean theory defined on the future boundary of spacetime.  The 
measure of the multiverse can then be related to the short-distance 
cutoff in that theory.  Related ideas have been explored in 
\cite{Bousso09,BFLR10}. 

Even without knowing the specific form of the boundary theory, one can 
try to deduce some properties of the resulting measure.   In particular, 
it has been argued in \cite{CAH} that in spacetime regions with a 
slowly varying expansion rate $H$, the corresponding cutoff surfaces are 
the surfaces of constant comoving apparent horizon (CAH).  Extension of 
the cutoff to regions where the variation of $H$ is not slow will 
require a better understanding of the boundary-bulk correspondence. In 
the meantime, it was suggested in \cite{CAH} that we could formulate a 
simple measure prescription which has no apparent pathologies and agrees 
with the CAH cutoff in regions of slow $H$ variation.  The hope is that 
such a prescription may be useful as a simple model of the measure until 
further progress is made. 

A guiding principle that can be used to extend the CAH cutoff is the 
causality condition, which requires that the cutoff surfaces should be 
spacelike, or in the limiting case null \cite{CAH}.  Otherwise the 
resulting measure could assign nonzero probabilities to some events in 
the absence of their causes. (An additional reason for imposing this 
condition in \cite{CAH} was that the holographic 
prescription adopted there identified the cutoff surfaces with the 
spacelike 3D surfaces on which the wave function of the universe is 
defined.)  We shall see in the next section that constant CAH surfaces 
generally include timelike segments, so they do not qualify as cutoff 
surfaces in the entire spacetime.  This problem can be fixed by 
introducing an additional simple rule which enforces causality.  It says 
that, given a constant CAH surface $\Sigma$, we should remove the 
spacetime region to the future of $\Sigma$.  As a result $\Sigma$ is 
replaced by a new cutoff surface $\Sigma_+$, which is the boundary of 
the future of $\Sigma$.\footnote{A similar modification (but with a 
different motivation) was suggested for the scale-factor cutoff measure 
in \cite{DeSimone:2008if}.}   If $\Sigma$ is spacelike, then $\Sigma_+$ 
and $\Sigma$ are the same, but if $\Sigma$ includes timelike segments, 
it will be modified in such a way that $\Sigma_+$ is not timelike 
anywhere.  The surface $\Sigma_+$ will generally include null segments, 
but these can be made spacelike by an infinitesimal deformation of the 
surface.  The corresponding measure was called the CAH+ measure in 
\cite{CAH}. 

This paper studies the phenomenological properties of the CAH+ 
measure, in particular its predictions for the cosmological constant 
and for the density parameter.  In the next section we review the 
definition of the apparent horizon and discuss some relevant properties 
of constant CAH surfaces.  The phenomenology of the CAH+ measure is 
discussed in Section \ref{sec:phenomenology}.  To put this analysis in 
a wider context, in Section \ref{sec:review} we review the key 
phenomenological properties of some other measure proposals.  Our 
conclusions are summarized and discussed in Section \ref{sec:discussion}.

\section{The CAH+ measure}
\label{sec:cah+}

\subsection{Definition}
\label{sec:definition}

We first need to define what is meant by the apparent horizon (AH) in a 
general spacetime.  For any spacelike 2D surface ${\mathcal S}$ we can 
construct two null hypersurfaces emanating orthogonally from 
${\mathcal S}$ to the future, one corresponding to outward and the other 
to inward directed light rays.  We shall call ${\mathcal S}$ an AH if 
the outward going null geodesics are expanding, while the inward going 
null geodesics have zero expansion (that is, they are neither diverging 
nor converging as they leave ${\mathcal S}$).\footnote{Note that this 
definition is different from that given by Bousso in \cite{Bousso02}.  
We define AH as a 2D surface, while Bousso defines it as a 3D 
hypersurface.  Also, his definition refers to a specific observer, and 
the AH surface depends on the entire observer's worldline.  In contrast, 
our definition depends only on the local geometry.}  
An apparent horizon defined in this way is a marginally anti-trapped 
surface.  

The next step is to define the CAH.  We start with a smooth 
segment of spacelike hypersurface $\Sigma_0$, located in an inflating 
region of some Hubble rate $H_0$ and having three-curvature
$|R^{(3)}|\ll H_0^2$.  We then construct a future-directed, timelike 
geodesic congruence orthogonal to $\Sigma_0$, labeling the geodesics 
by their starting points ${\bf x}$ on $\Sigma_0$.  The scale factor 
$a({\bf x},\tau)$ can be defined as the cubic root of the volume 
expansion factor along the geodesic at ${\bf x}$ in a proper time 
$\tau$, with $\tau=0$ and $a({\bf x},0)=1$ on $\Sigma_0$.  The 
expansion rate of the congruence is
\beq
H({\bf x},\tau)={\dot a}({\bf x},\tau)/a({\bf x},\tau),
\eeq
where dots stand for derivatives with respect to $\tau$. 

Let us first assume that the spacetime can be locally approximated as 
FRW, with our geodesic congruence playing the role of comoving 
geodesics.  This should be a good approximation in inflating regions 
away from bubble walls and in thermalized regions, as long as effects 
of structure formation can be neglected (we expect such effects to be 
small on the scale of the AH).  In this case the AH surfaces 
lying in three-spaces orthogonal to the congruence are spheres 
of radius, $r_{\rm AH} = H^{-1}({\bf x},\tau)$.  The comoving apparent 
horizon radius is then 
\beq
r_{\rm CAH}({\bf x},\tau)=\big[a({\bf x},\tau)H({\bf x},\tau)\big]^{-1}
={\dot a}^{-1}({\bf x},\tau) \,.
\label{rCAH}
\eeq
It is convenient to define the cutoff foliation in terms of the inverse 
of this quantity, the ``CAH time''
\beq
\theta\equiv r_{\rm CAH}^{-1} = \dot{a} \,.
\eeq

In general, the AH surfaces will not be spherical, and this 
leads to an ambiguity in the definition of the CAH.  We could, 
for example, define a constant CAH hypersurface $\Sigma$ by requiring 
that all AH surfaces on $\Sigma$ enclose the same volume or have the 
same maximal extent when projected along the geodesic congruence onto 
the hypersurface $\Sigma_0$.  At the level of our present understanding 
we cannot give preference to any of the alternative definitions, but we 
do not expect these differences to significantly affect the measure.  We 
shall therefore choose the simplest option and use the prescription of 
(\ref{rCAH}) in what follows.  The quantity $r_{\rm CAH}$ in (\ref{rCAH}) 
can be interpreted as the average CAH radius.

\subsection{Practical application and approximations}
\label{sec:approximation}

In the calculations below we adopt a number of simplifying assumptions 
and approximations, which we now discuss.

We consider our universe to be part of a Coleman--De Luccia (CDL) 
bubble \cite{CDL}, nucleating in an eternally-inflating parent vacuum 
characterized by the Hubble rate $\Hp$.  In this context an additional 
period of (slow-roll) inflation within the bubble is necessary, to 
redshift the initial spatial curvature of the bubble to observationally 
acceptable levels.  We approximate this as $N$ $e$-folds of 
exponential expansion with Hubble rate $\approx\Hd$.  

To be precise, CDL bubble nucleation generates a bubble geometry 
of the form
\beq
ds^2 = -d\tau^2 
+ \a^2(\tau)\Big[d\xi^2+\sinh^2(\xi)\,d\Omega_2^2\Big] \,,
\label{FRW0}
\eeq
where $d\Omega_2$ is the infinitesimal line element on the unit 
two-sphere, and we use the symbol $\a$ for the scale factor in
the bubble, to distinguish it from the more global notion 
referred to above.  (We everywhere assume 3+1 spacetime 
dimensions.)  A scale-factor solution corresponding to 
inflation characterized by asymptotic Hubble rate $\Hd$ (after 
an initial period of curvature domination, which is necessary to 
match to the CDL instanton boundary conditions) is
\beq
\a(\tau) = \Hd^{-1}\,\sinh(\Hd\tau) \,.
\label{a0}
\eeq
We assume inflation ends (abruptly) at $\tau_\star=N\Hd^{-1}$.  
This makes precise what is meant by the inflationary Hubble rate 
$\Hd$ and the number of $e$-folds $N$.  After inflation we assume
instantaneous reheating, followed by radiation domination, 
non-relativistic matter domination, and then either spatial-curvature 
domination or cosmological constant domination, consistent with the 
observed big-bang evolution, except allowing for uncertainty as to 
the size of $\Lambda$ (and thus allowing for curvature 
domination preceding $\Lambda$-domination).  

To determine the location of the CAH cutoff, we need to track the 
evolution of the CAH time $\theta$ along a congruence of timelike 
geodesics that begin in the parent vacuum and enter the bubble.  The 
details of the calculation are a bit complicated, and have been 
relegated to Appendix \ref{sec:CDL}.  For simplicity we assume that 
the bubble radius at nucleation is much smaller than the Hubble radius 
$\Hp^{-1}$.  We also disregard the effect of the bubble wall on the 
geodesics that pass through it.  Otherwise our analysis is general, 
but the results can be further simplified if we assume the vacuum 
energy in the parent vacuum to be significantly larger than the 
inflationary energy density in the bubble, $\Hp\gg\Hd$.  We shall adopt 
this simplification below (the results are qualitatively unchanged for 
$\Hd\sim\Hp$).

\subsection{CAH and CAH+ cutoff surfaces}
\label{sec:surfaces}

The CAH time 
in the bubble must be defined with respect to some ``initial'' 
constant-CAH-time hypersurface $\Sigma_0$, which we momentarily 
place in the parent vacuum.  Then, at times $\tau\gg\Hd^{-1}$ 
during inflation in the bubble, the CAH time is   
\beq
\theta_{\rm inf}(\tau,\,\xi) = \frac{1}{2}\,\theta_{\rm n}\, 
e^{\xi+\Hd\tau} \,,
\label{t1}
\eeq
where the factor $\theta_{\rm n}$ corresponds to the CAH time 
at the point of bubble nucleation.  

To solve for the scale-factor evolution after inflation, we assume
instantaneous transitions between radiation domination beginning 
at $\tau=\tau_\star$, non-relativistic matter domination beginning at
$\tau=\tau_{\rm eq}$, spatial-curvature domination beginning at 
$\tau=\tc$, and/or cosmological-constant domination 
beginning at $\tau=\tau_\Lambda$, matching the scale factor and its
first derivative at each transition.  (The details are presented in
Appendix \ref{sec:bigbang}.)  The CAH time is likewise obtained by 
matching; thus the CAH time after inflation is given by substitution
into
\beq
\theta(\tau,\,\xi)= \theta_{\rm inf}(\tau_\star,\,\xi)\, 
\frac{\dot{\a}(\tau)}{\dot{\a}(\tau_\star)} \,.
\label{t2}
\eeq
For example, during radiation domination we have
\beq
\theta_{\rm rad}(\tau,\,\xi) = \frac{1}{2}\,\theta_{\rm n}\,
e^{\xi+N} \big(2\Hd\tau-2N+1\big)^{-1/2} \,.
\label{t3}
\eeq

From the definition of $\theta$ it follows that
\beq
{\dot\theta} = {\ddot a},
\eeq
so the CAH time grows with proper time along the geodesics in the 
inflating regions of spacetime, where ${\ddot a}>0$.  On the other 
hand, in thermalized regions, which include radiation, matter, and 
curvature dominated epochs inside the bubbles, ${\ddot a}<0$ and 
$\theta$ decreases along the geodesics.  Thus, if a comoving geodesic 
reaches a thermalized region before reaching the cutoff CAH time 
$\theta_{\rm cut}$, it will traverse the entire thermalized region 
without reaching $\theta_{\rm cut}$.  For Minkowski bubbles 
$(\Lambda=0)$ and AdS bubbles ($\Lambda< 0$), CAH time decreases all 
the way up to future infinity / the big crunch singularity (in the 
former case $\theta$ asymptotically approaches a constant).  Only 
in bubbles with $\Lambda>0$ does $\theta$ again increase with FRW 
time, when $\Lambda$ comes to dominate the universe.

\begin{figure}[t!]
\begin{center}
\includegraphics[width=0.4\textwidth]{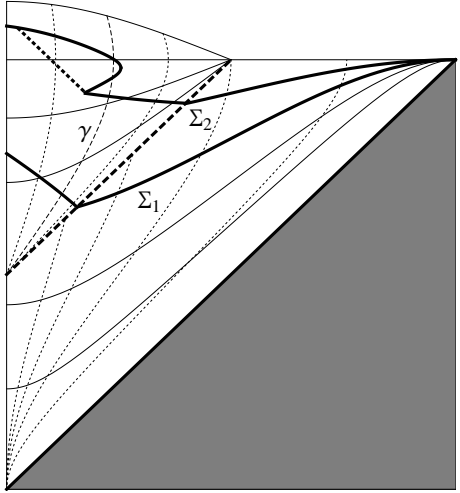}
\caption{\label{fig:cutoff} Two constant--CAH time hypersurfaces, 
$\Sigma_1$ and $\Sigma_2$ (thick, solid curves), intersecting 
and extending into a bubble universe.  The thin, solid curves 
correspond to constant-$t$ (constant-$\tau)$ surfaces, while the
thin, dotted curves correspond to constant-$r$ (constant-$\xi$) 
surfaces, in the parent vacuum (bubble).  The thick, dashed 
line approximates the bubble wall, while the thick, dotted 
line represents the cutoff imposed by the CAH+ 
measure prescription.  A comoving FRW geodesic $\gamma$ is 
indicated by the dashed line. (See main text.)}
\end{center}
\end{figure}

To illustrate these dynamics, in Figure \ref{fig:cutoff} we 
project two constant-$\theta$ hypersurfaces onto a conformal 
diagram of positive-$\Lambda$ CDL bubble nucleation (the 
spatially-flat de Sitter chart of the parent vacuum covers only the 
upper-left half of the diagram).  In this diagram, $\Sigma_1$ 
represents a CAH time that does not
probe beyond the inflationary epoch in the bubble, and is 
therefore entirely described by (\ref{t1}) in the bubble.  As
the CAH time $\theta$ is increased, the curve migrates to larger 
FRW times, and its image on the diagram flattens out.  For 
sufficiently large $\theta$ and sufficiently small $\xi$, the curve 
probes beyond the inflationary epoch in the bubble.  Then, since 
$\theta(\tau,\,\xi)$ decreases with FRW time $\tau$ along constant- 
and decreasing-$\xi$ trajectories, the curve runs toward increasing 
values of $\xi$ as $\tau$ is increased.  This persists until 
cosmological-constant domination, when $\theta$ again increases 
with $\tau$ along constant-$\xi$ trajectories, and the curve runs to 
smaller $\xi$ for increasing $\tau$, until it reaches $\xi=0$.  The 
curve $\Sigma_2$ represents a cartoon of such dynamics.  

Consider for example the FRW geodesic labeled $\gamma$ in Figure 
\ref{fig:cutoff}.  This geodesic reaches the cutoff value 
$\theta_{\rm cut}$ on the surface $\Sigma_2$ while it is still in the 
inflating region.  Beyond this point, $\theta$ continues to grow 
until the end of inflation.  Then it decreases until the value 
$\theta_{\rm cut}$ is reached again.  This is where the geodesic 
crosses $\Sigma_2$ for the second time.  After this crossing, 
$\theta$ keeps decreasing until the onset of $\Lambda$ domination, 
where it starts to grow again, and finally reaches $\theta_{\rm cut}$ 
for the third time, at the third crossing of $\Sigma_2$.  If the 
cosmological constant had been precisely zero, the CAH 
time $\theta$ would never have resumed increasing with FRW time, and
$\Sigma_2$ would continue to run to larger values of $\xi$ as $\tau$ 
is increased, until reaching spacelike infinity.  The geodesic $\gamma$ 
would then cross $\Sigma_2$ only twice, while FRW geodesics at 
sufficiently small values of $\xi$ would never cross $\Sigma_2$.  The 
same qualitative picture applies for negative $\Lambda$.

Now, the CAH+ prescription requires that we remove all points in the 
future light cone of any point on the constant-CAH surface $\Sigma$.  
This leaves $\Sigma_1$ unchanged, but it does modify $\Sigma_2$, 
replacing a part of it with a null hypersurface, as indicated by the 
thick, dotted line in Figure \ref{fig:cutoff}.

So far we have discussed the constant CAH-time foliation in a CDL 
bubble, which is defined with respect to a spatially flat initial 
hypersurface $\Sigma_0$ in the parent vacuum.  One might worry about 
the generality of this setup.  However, during exponential expansion, 
timelike geodesics rapidly converge to comoving worldlines in the 
spatially-flat foliation.  Furthermore, at late times in a positive 
vacuum-energy bubble, the CAH-time foliation asymptotes to that of a 
spatially-flat de Sitter chart (see Appendix \ref{sec:CDL}).  This 
means it is not important for the hypersurface $\Sigma_0$ to be flat 
or to have been drawn in the parent vacuum, as opposed to in any 
previous ancestor vacuum.  Any spacelike hypersurface $\Sigma_0$ would 
do, as long as it is placed sufficiently deep in the past of the 
bubble nucleation event.

\subsection{Prior distribution}
\label{sec:prior}

The bubble nucleation time $\theta_{\rm n}$, although a constant with 
respect to the evolution of $\theta(\tau,\,\xi)$ in any given bubble, 
does depend on the history of the subset of the congruence entering 
the bubble.  The statistical properties of $\theta_{\rm n}$ are 
determined by the coarse-grain rate equation, which tracks the 
three-volume ${\cal V}_i$ occupied by a given vacuum $i$ on surfaces of
constant CAH time $\theta$, averaging over (proper) timescales that are
large compared to the relevant Hubble times, but small compared to the 
the relevant inverse decay rates.\footnote{We thank Daniel Harlow for
pointing out a mistake in a previous analysis.}

It is convenient to start with the rate equation in terms of
scale-factor time $t$ \cite{Garriga:1997ef,DeSimone:2008if},
\beq
\frac{d{\cal V}_i(t)}{dt} = 3{\cal V}_i(t) 
+ \sum_j \kappa_{ij} {\cal V}_j(t) - \sum_j \kappa_{ji} {\cal V}_i(t) \,,
\label{sfrate}
\eeq
where $\kappa_{ij}\equiv (4\pi/3)H_j^{-4}\Gamma_{ij}$ is the 
dimensionless decay rate.  This equation is understood to apply
only to de Sitter vacua $i$, and the first term on the right-hand-side
accounts for the volume expansion during positive vacuum energy 
domination.  The second the second term sums over bubble nucleations of 
vacuum $i$ in other vacua $j$, and the last term sums over decays of 
vacuum $i$ to vacua $j$.  The last sum is understood to run over de 
Sitter and Anti--de Sitter (and Minkowski) vacua (the first sum can 
also be taken to run over all vacua, however only de Sitter vacua
contribute since the other transition rates are zero).    

To convert to CAH time, first note that the difference between 
two scale-factor times evaluated at two points ``1'' and ``2'' along 
a given geodesic can be written
\beq
t_2-t_1 = \ln(a_2/a_1) = \ln(\theta_2/\theta_1) - \ln(H_2/H_1) \,,
\label{sfcah}
\eeq 
where we have used $t = H\tau = \ln(a) = \ln(\theta/H)$ in the 
spatially-flat slicing of de Sitter space, and the subscripts denote 
at which point a given quantity is evaluated.  It can be shown that 
this relationship between scale-factor and CAH times holds not only 
within a single bubble (where $H_2=H_1$), but also along a geodesic 
traversing two or more bubbles, so long as the proper time between
points ``1'' and ``2'' is much larger than the Hubble time.      

We have made the scale-factor time dependence of (\ref{sfrate}) 
explicit to make clear that converting to CAH time involves not only
changing the differential time element,
\beq
dt \,\to\, d\ln(\theta/\theta_0) \,,
\eeq    
with $\theta_0$ here being an arbitrary scale, but also changing the 
volume ${\cal V}_j(t)$, since surfaces of constant scale-factor time
are not surfaces of constant CAH time, in the presence of bubble
nucleations.  In particular, the volume expansion factor in terms of
scale-factor time is $e^{3t}$, and therefore the logarithmic offset 
of (\ref{sfcah}) corresponds to a shift in time with concomitant 
volume expansion factor $(H_1/H_2)^3\to (H_j/H_i)^3$ multiplying
${\cal V}_j$.  Thus,
\beq
\frac{d{\cal V}_i(\theta)}{d\ln(\theta/\theta_0)} = 3{\cal V}_i(\theta) 
+ \sum_j \kappa_{ij}\,(H_j/H_i)^3\,{\cal V}_j(\theta) 
- \sum_j \kappa_{ji} {\cal V}_i(\theta) \,.
\label{cahrate}
\eeq
Note that this is also the rate equation for the lightcone time measure 
of \cite{Bousso09,BFLR10}, since lightcone time and CAH time are 
equivalent within de Sitter vacua.   

It is convenient to define the quantity
\beq
f_i \equiv H_i^3{\cal V}_i \,,
\label{fidef}
\eeq
so that the rate equation can be written
\beq
\frac{df_i}{d\ln(\theta/\theta_0)} = 3f_i + 
\sum_j \kappa_{ij} f_j -
\sum_j \kappa_{ji} f_i \,.
\label{rateeq}
\eeq
The equation is now a simple change in variables from that studied in 
\cite{Garriga:1997ef,DeSimone:2008if}.  In particular, at late times 
the solution can be written
\beq
f_i(\theta) = s_i\,(\theta/\theta_0)^{3-q} + \ldots \,,
\label{ratesol}
\eeq
where $q>0$ is the smallest-magnitude eigenvalue of the transition
matrix $\kappa_{ij}-\delta_{ij}\sum_k\kappa_{ki}$, $s_i$ is the 
corresponding eigenvector, and the ellipses denote terms that fall 
off faster than $\theta^{-q}$.  Using (\ref{fidef}) to solve for 
${\cal V}_i$, we find
\beq
{\cal V}_i(\theta) \propto s_i\,(\theta_0/H_i)^q\,(\theta/H_i)^{3-q}\,,
\eeq     

Therefore, the number of bubble nucleations of type $i$ in
a CAH time interval $d\theta$ is
\beq
d{\cal N}_i(\theta) = \sum_j \Gamma_{ij}\,{\cal V}_j(\theta)\,
\frac{d\tau}{d\theta}\bigg|_j d\theta \,\propto\, 
\sum_j \kappa_{ij}s_j\, \theta^{2-q}\,d\theta \,.
\label{dNi}
\eeq
(The factor  $d\tau/d\theta$ in the first expression arises because
the rate $\Gamma_{ij}$ is given per unit proper time in the vacuum 
$j$.)  In practice, the exponent $q$ is on the order of the smallest 
(dimensionless) decay rate among positive-energy vacua in the 
landscape, and is therefore negligible next to the power of two.  The 
factor
\beq
P_i \propto \sum_j \kappa_{ij}s_j
\label{Pi}
\eeq
gives the relative number of bubbles of type $i$ below the cutoff.  
It can be regarded as the ``prior'' probability for this type of bubble.

\section{Phenomenology}
\label{sec:phenomenology}

Section \ref{sec:approximation} describes evolution of CAH time 
$\theta$ as a function of the coordinates $(\tau,\,\xi)$ in a CDL 
bubble.  To make predictions, we set a cutoff value $\theta_{\rm cut}$, 
and compute statistics according to the relative numbers of different 
types of events between the initial hypersurface $\Sigma_0$ and the 
cutoff hypersurface $\Sigma_{\rm cut}$.  We focus on the subset of 
bubbles that are indistinguishable from ours, except for the value of 
the cosmological constant $\Lambda$ and the number of inflationary 
$e$-folds $N$.  The events of interest to us here are the observations 
of $\Lambda$ and $N$, so they can be labeled by the same index $i$ as 
the bubbles.  The number of such events in bubbles of type $i$ between 
$\Sigma_0$ and $\Sigma_{\rm cut}$ can be written as
\beq
{\cal N}_i \propto P_i \int_{\theta_0}^{\theta_{\rm cut}} 
d\theta_{\rm n}\, \theta_{\rm n}^{2-q}
\int_0^{\tau_{\rm cut}}
d\tau\,\a^3(\tau)\,\rho_i(\tau)
\int_0^{\xi_{\rm cut}}
d\xi\,\sinh^2(\xi) \,.
\label{int}
\eeq
Here $P_i$ is the prior probability (\ref{Pi}) for bubbles of type $i$, 
$\rho_i(\tau)$ is the number of relevant observations per unit physical 
four-volume in such bubbles, 
$\xi_{\rm cut}(\tau,\,\theta_{\rm n},\,\theta_{\rm cut})$ is the 
value of $\xi$ at which the constant-$\tau$ hypersurface intersects 
the cutoff hypersurface, and
$\tau_{\rm cut}(\theta_{\rm n},\,\theta_{\rm cut})$ is the maximum 
FRW proper time in the bubble probed by the cutoff hypersurface.  
Moving from right to left, the first two integrations count the 
number of observations under the cutoff in a bubble of type $i$ 
nucleating at CAH time $\theta_{\rm n}$, while the last integral 
sums over bubble nucleation times, according to Eq.(\ref{dNi}).
Since $q$ is exponentially small, we henceforth drop it.  

Since the CAH+ measure prescription involves augmenting the 
cutoff hypersurface with lightcones, it is convenient to work in
terms of the bubble conformal time $\eta=\int d\tau/\a(\tau)$, 
as opposed to the proper time $\tau$.  During the early-time 
inflationary epoch in the bubble, i.e.~for 
$|\eta_\star| \ll |\eta| \ll 1$, where $\eta_\star=-2e^{-N}$ is 
the reheating time, CAH time is given by
\beq
\theta(\eta,\,\xi)=-\frac{\theta_{\rm n}}{\eta}\,
e^{\xi}\,.
\label{theta1}
\eeq
After inflation, (\ref{theta1}) no longer holds. (We 
assume instantaneous reheating.)  As described in Section 
\ref{sec:approximation}, during radiation, non-relativistic matter, 
and spatial-curvature domination, $\theta$ decreases along 
comoving geodesics.  The would-be cutoff hypersurface 
$\theta=\theta_{\rm cut}$ therefore runs toward increasing 
$\xi$ with increasing $\eta$, and in the CAH+ prescription it is 
augmented by the future lightcone of where it intersects the 
reheating hypersurface, $\theta = (1/2)\,\theta_{\rm n}\,e^{\xi+N}$.  
When $\eta>\eta_\star$, this lightcone provides the cutoff
\beq
\xi_{\rm cut} = - \eta + \eta_\star +
\ln(2\theta_{\rm cut}/\theta_{\rm n})-N \,.
\label{xi1}
\eeq
The events that concern us (observers like us measuring 
cosmological parameters) occur only after non-relativistic matter 
domination, when $\eta\gg\eta_\star$, so henceforth we drop 
$\eta_\star$ next to $\eta$.

In bubbles with zero or negative cosmological constant, the 
cutoff (\ref{xi1}) applies to all times after reheating.  In bubbles
with positive cosmological constant, however, $\theta$ begins to 
increase again at cosmological-constant domination, and the 
$\theta=\theta_{\rm cut}$ hypersurface can supercede (\ref{xi1}).
The evolution of $\theta$ after cosmological-constant domination 
is determined by (\ref{t2}), substituting conformal time 
$\eta$ for proper time $\tau$ (see Appendix \ref{sec:bigbang}).  
The result depends on whether or not there is a period of 
spatial-curvature domination after non-relativistic matter 
domination, in particular:
\bea
\theta = 
\left\{ \,\,
\begin{array}{ll}
\displaystyle 
\frac{\theta_{\rm n}\,e^{\xi}}
{3-\ln(\textstyle\frac{3}{2}\HL\tc)-\eta} 
\quad\!\! & \phantom{\Bigg[\Bigg]} 
\mbox{if there is curvature domination}\,\,\,\, \\
\displaystyle
\frac{\theta_{\rm n}\,e^{\xi}}
{(\textstyle\frac{1}{18}\HL\tc)^{-1/3}-\eta}
\quad\!\! & \phantom{\Bigg[\Bigg]}
\mbox{if not} \,,
\end{array} \right.
\label{e2a}
\eea
where $\HL\equiv\sqrt{|\Lambda|/3}$, and $\tau_{\rm c}$ is
the time at which spatial curvature would begin to dominate, if 
$\HL$ were zero.  In each case the expression is valid after the 
onset of cosmological-constant domination at $\eta_\Lambda$, 
corresponding to when the denominator is equal to unity.  

When (positive) cosmological-constant domination follows a 
period of spatial-curvature domination, the hypersurface 
$\theta=\theta_{\rm cut}$ is obtained by inverting the upper of 
(\ref{e2a}), and gives
\beq
\xi_{\rm cut} = \ln\!\left[3-\ln\!\left(\textstyle
\frac{3}{2}\HL\tc\right)-\eta\right]
+ \ln(\theta_{\rm cut}/\theta_{\rm n}) \,.
\label{xi2}
\eeq 
This is the actual CAH+ cutoff when it provides a stronger
constraint than the future lightcone cutoff (\ref{xi1}), so the
cutoff is given by the smaller of (\ref{xi1}) and (\ref{xi2}).  
Since both of these expressions are 
monotonically decreasing, (\ref{xi1}) provides the cutoff up 
to some time $\eta_1$, after which (\ref{xi2}) provides the 
cutoff, with $\eta_1$ corresponding to the solution of
\beq
2e^{-\eta_1-N} = 3-\ln\!\left(
\textstyle\frac{3}{2}\HL\tc\right)-\eta_1 \,. \label{e1}
\eeq 
A similar story holds when non-relativistic matter domination 
gives way directly to (positive) cosmological-constant domination.  
In this case the $\theta=\theta_{\rm cut}$ hypersurface gives
\beq
\xi_{\rm cut} = \ln\!
\left[\left(\textstyle\frac{1}{18}\HL\tc\right)^{-1/3}
-\eta\right] + \ln(\theta_{\rm cut}/\theta_{\rm n})\,.
\label{xi3}
\eeq
As before, the actual CAH+ cutoff is given simply by the smaller 
of (\ref{xi1}) and (\ref{xi3}).  Likewise, (\ref{xi1}) provides 
the cutoff up to some time $\eta_2$, after which (\ref{xi3}) 
provides the cutoff, with $\eta_2$ corresponding to the solution 
of
\beq
2e^{-\eta_2-N} = \left(
\textstyle\frac{1}{18}\HL\tc\right)^{-1/3}-\eta_2\,. \label{e2}
\eeq  
Henceforth we refer to the above transition times as $\eta_{1,2}$,
with it being understood that $\eta_1$ applies to when 
non-relativistic matter domination gives way directly to 
cosmological constant domination (which corresponds to 
$\tc\geq (2/3)\HL^{-1}$), and $\eta_2$ applies when there is a 
period of late-time spatial-curvature domination (corresponding
to $\tc< (2/3)\HL^{-1}$).

Putting the above results together, it is convenient to write
\beq
\xi_{\rm cut}(\eta,\theta_{\rm n}) = -f(\eta) + 
\ln(\theta_{\rm cut}/\theta_{\rm n}) \,,
\eeq
with $f(\eta)$ given by
\bea
f(\eta) = 
\left\{\,\,
\begin{array}{ll}
\displaystyle 
\eta+N-\ln(2) \phantom{\bigg[\bigg]}\,\,\, & 
\parbox{205pt}{
if $\Lambda\leq 0$, or \\
if $\Lambda > 0$ and $\eta<\eta_{1,2}$}\\
\displaystyle
\ln\!\big[\HL\a(\eta)\big]
\phantom{\bigg[\bigg]}\,\,\, & 
\parbox{200pt}{
if $\Lambda > 0$ and $\eta\geq\eta_{1,2}$\,,}\\
\end{array} \right.
\label{f}
\eea
where we have used the results of Appendix \ref{sec:bigbang}
to recognize the dependence on $\a$.  We are now prepared to return 
to (\ref{int}).  Performing the integration over $\xi$, we obtain
\beq
{\cal N}_i \propto P_i \int_{\theta_0}^{\theta_{\rm cut}} 
d\theta_{\rm n}\, \theta_{\rm n}^2 
\int_{\eta_\star}^{f^{-1}[\ln(\theta_{\rm cut}/\theta_{\rm n})]}\! 
d\eta\,\,\a^4(\eta)\,\rho_i(\eta)\, 
\Big\{\sinh\!\big[2\xi_{\rm cut}(\eta,\theta_{\rm n})\big]
-2\xi_{\rm cut}(\eta,\theta_{\rm n}) \Big\}\,. \,\,
\label{int2}
\eeq   
It is possible to exchange the order of $\theta_{\rm n}$ and 
$\eta$ integration.  It is also worthwhile to change the 
integration variable $\theta_{\rm n}$ to 
$y\equiv\ln(\theta_{\rm cut}/\theta_{\rm n})-f(\eta)$.  These
operations allow us to write
\beq
{\cal N}_i \propto P_i \int_{\eta_\star}^{\eta_{\rm max}}\!
d\eta\,\,\a^4(\eta)\,\rho_i(\eta)\, e^{-3f(\eta)} \,
{\cal I}(\eta) \,, \label{Ni}
\eeq
where $\eta_{\rm max} \equiv \min\{\,
\ln(\theta_{\rm cut}/\theta_0), \mbox{the maximum value of $\eta$ 
reached in the bubble}\}$, and
\beq
{\cal I}(\eta)\equiv \int_0^{\ln(\theta_{\rm cut}/\theta_0)-f(\eta)} 
dy\, e^{-3y}\Big[\sinh(2y)-2y\Big] \,.
\eeq
In the limit $\theta_{\rm cut}\to\infty$, assuming that 
$\rho_i(\eta)$ is localized within some finite range of $\eta$, 
we can take the upper limit of integration of ${\cal I}(\eta)$ 
to infinity, so that ${\cal I}(\eta)$ becomes a constant.

\subsection{Geometric effects}
\label{sec:geometry}

Before focusing on the implications of (\ref{Ni}) for observers 
like us, it is worthwhile to investigate some of the more general 
features of this result.  Toward this end we here adopt a more 
crude ``anthropic'' model, which places all observers at 
some fixed FRW proper time $\tobs$ in the bubble, and takes the 
density of their observations to be proportional to the density 
of non-relativistic matter.  (Our analysis follows the spirit of 
\cite{BFLR}.)  Intuitively, it would be preferable to place
the observers at some fixed proper time after reheating, but we 
assume $\tobs\gg N\Hd^{-1}$ so that the difference is 
negligible.  To be precise, we write
\beq
\rho_{\rm obs}(\eta) \propto \frac{\a^3(\eta_\star)}{\a^4(\eta)}\,
\delta\big[\eta-\eta(\tobs)\big] \,,
\eeq
where $\delta$ is the Dirac delta function, and we have included 
a factor of $|d\tau/d\eta|^{-1}=\a^{-1}$ to account for the 
measure of integration, as well as a factor of $\a^3(\eta_\star)$ 
because inflationary expansion does not dilute the density of 
non-relativistic matter after reheating.  This gives
\beq
{\cal N}_{\rm obs} \propto P(\Lambda,N)\,
e^{3N-3f[\eta(\tobs)]}\,,
\label{Nobs1}
\eeq
where the ``prior'' $P(\Lambda,N)$ is the probability that a random 
bubble will be characterized by given values of $N$ and $\Lambda$ 
and be otherwise identical to ours.  

It is convenient to express $\Lambda$ and $\tc(N)$ in units 
of $\tobs$.  Accordingly, we write
\beq
\ell = {\rm sign}(\Lambda)\,(\HL\tobs)^2 \,,
\qquad {\rm and} \qquad
\tc=\tobs\,e^{3(N-N_{\rm obs})}\,,
\eeq 
where $N_{\rm obs}$ depends on the details of the cosmological 
history.  Then $\ell\approx 1$ corresponds to observers arising at 
about the onset of cosmological-constant domination, and 
$N= N_{\rm obs}$ corresponds to observers arising at the onset of 
spatial-curvature domination (in bubbles where this happens).  If 
we consider $\ell$ and $\tc$ to be fixed and study 
${\cal N}_{\rm obs}$ as a distribution over $\tobs$, then we 
find the distribution features two qualitatively distinct regimes.  
For positive $\ell$ and sufficiently large $\tobs$, we have
$\eta(\tobs)\geq\eta_{1,2}$ and 
${\cal N}_{\rm obs}\propto e^{3N}\big[\HL\a(\tobs)\big]^{-3}$.  
Thus ${\cal N}_{\rm obs}$ is proportional to the density of 
non-relativistic matter.  For smaller 
$\tobs$, as well as for $\ell\leq 0$, 
${\cal N}_{\rm obs}\propto e^{-3\eta(\tobs)}$.  This corresponds
to a weaker (decreasing) function of $\tobs$ than the density of 
non-relativistic matter.  Thus we conclude that 
${\cal N}_{\rm obs}$ prefers smaller values of $\tobs$, but the
effect is not stronger than the youngness bias of the scale-factor
cutoff or fat geodesic measures \cite{DSGSV,Bousso:2008hz}.  In 
particular, if we were to condition on structure formation, 
${\cal N}_{\rm obs}$ would become strongly suppressed for 
sufficiently small $\tobs$, resulting in a localized distribution.  
To simplify the discussion, for the rest of this subsection we assume 
$\tobs$ is fixed, and corresponds to some time during or after 
non-relativistic matter domination.

Since the CAH+ measure does not exponentially favor large 
amounts of inflation in the bubble, it does not suffer from the $Q$ 
and $G$ catastrophes \cite{Feldstein:2005bm,Garriga:2005ee,
Graesser:2006ft}.  Note that the CAH+ measure can also avoid 
Boltzmann-brain domination \cite{Dyson:2002pf,Albrecht:2004ke,
Page:2006dt}.  In particular, we found above that for sufficiently 
late times, $\eta(\tobs)\geq\eta_{1,2}$, the measure samples in 
proportion to $a(\tobs)^{-3}$.  Modulo proportionality constants, 
this is the same late-time behavior as the causal patch, scale-factor 
cutoff, and fat geodesic measures, each of which has been shown to 
be able to avoid Boltzmann-brain domination \cite{Bousso:2006xc,
DeSimone:2008if,Bousso:2008hz}.  As is discussed in those papers, 
the early-time behavior is unimportant compared to the huge 
timescales on which Boltzmann brains typically form.  Similarly the
differing proportionality constants are inconsequential next to 
Boltzmann-brain nucleation rates and vacuum decay rates that 
dominate the calculations.

Consider now ${\cal N}_{\rm obs}$ as a distribution over $N$, 
keeping $\ell$ (and $\tobs$) fixed.  For $N<N_{\rm obs}$ and 
$|\ell|<2/3$, the time $\tobs$ corresponds to during 
spatial-curvature domination.  In this regime, 
$\eta(\tobs)<\eta_{1,2}$, so we have 
${\cal N}_{\rm obs}\propto e^{-3\eta(\tobs)}$.  Using 
(\ref{acurveta}), we find 
${\cal N}_{\rm obs}\propto \big(\frac{2}{3}e^{3(N_{\rm obs}-N)}
+\frac{1}{3}\big)^{-3}$, i.e.~the distribution is exponentially
suppressed for $N$ significantly less than $N_{\rm obs}$. 
Surveying increasing values of $N$, $\tobs$ corresponds to during 
non-relativistic matter domination and ${\cal N}_{\rm obs}/P$ 
asymptotes to a constant.  The transition from exponentially 
increasing in $N$ to independent of $N$ occurs rather sharply at 
$N\sim N_{\rm obs}$; thus, for prior distributions $P$ that are 
(negative) power laws in $N$, the distribution ${\cal N}_{\rm obs}$ 
is peaked at $N\sim N_{\rm obs}$, but can have a long tail toward
large $N$.  The above discussion applies only to $|\ell|<2/3$, 
yet for larger values of $|\ell|$ the story is mostly the same: 
for a given value of $N$ the time $\tobs$ may correspond to during 
cosmological-constant domination instead of during spatial-curvature 
or non-relativistic matter domination, but overall the dependence of 
${\cal N}_{\rm obs}/P$ on $N$ is qualitatively unchanged from above.  
The only significant development occurs when $\ell$ becomes so large 
that $\eta(\tobs)>\eta_{1,2}$, in which case ${\cal N}_{\rm obs}/P$ 
becomes strictly independent of $N$.  As indicated in the next 
paragraph, ${\cal N}_{\rm obs}/P$ is suppressed in this region of 
the parameter space.

Finally, we discuss the dependence of ${\cal N}_{\rm obs}$
on $\ell$, keeping $N$ and $\tobs$ fixed.  Starting at 
sufficiently small values of $|\ell|$, the time $\tobs$ corresponds 
to during non-relativistic matter or spatial-curvature domination, 
and ${\cal N}_{\rm obs}/P$ is independent of $\ell$.  Surveying
increasing values of $|\ell|$, eventually $\tobs$ corresponds to 
during cosmological-constant domination.  When $\ell<0$, 
${\cal N}_{\rm obs}/P$ is a weakly decreasing function of $|\ell|$, 
until the minimum-allowed value of $\ell$ is reached when $\tobs$ 
coincides with the big-crunch singularity at $\ell\sim -(2\pi/3)^2$, 
the precise value of the bound depending weakly on $N$.  When 
$\ell>0$ the distribution ${\cal N}_{\rm obs}/P$ is a weakly 
increasing function of $\ell$, until $\eta(\tobs)$ becomes greater 
than $\eta_{1,2}$, after which the distribution decreases 
exponentially in $\ell$.  It can be shown that this transition 
occurs roughly at $\ell^{1/2}+\ln(\ell^{1/2})\sim 3N_{\rm obs}-2N$, 
when $N e^{3(N-N_{\rm obs})}\lesssim 1$, or at 
$\ell^{1/2}+(1/3)\ln(\ell^{1/2})\sim N_{\rm obs}$, when
$N e^{3(N-N_{\rm obs})}\gtrsim 1$.  If we neglect the 
logarithms and use $N\sim N_{\rm obs}$ from the previous paragraph, 
we see that the scale of $\ell$ is set by $N_{\rm obs}^2$.  In 
bubbles otherwise like ours, $N_{\rm obs}$ can be as large as $60$ 
(for GUT-scale inflation), setting the scale at a few thousand 
times the value of the cosmological constant we observe.  As we 
shall see, however, further consideration of anthropic selection
effects for observers like us can suppress such large values of 
$\ell$.

\begin{figure}[t!]
\begin{center}
\includegraphics[width=0.5\textwidth]{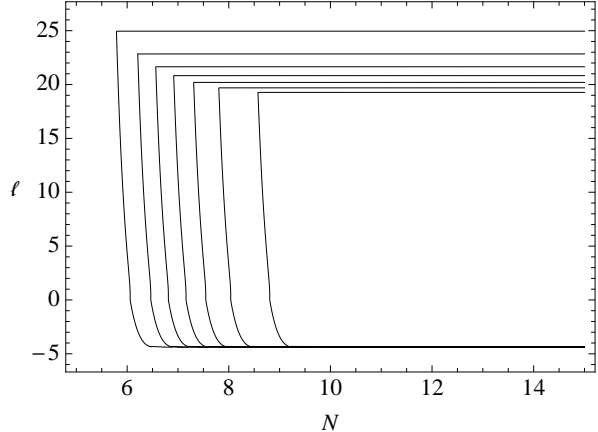}
\caption{\label{fig:NL0} Contour plot of the distribution 
${\cal N}_{\rm obs}/P$ in the $(N,\ell)$ plane (the distribution 
peaks toward the right and center).  We have set $N_{\rm obs}=5$ 
to condense the scale of the plot.}
\end{center}
\end{figure}

These results are illustrated in Figure \ref{fig:NL0}.  
We have chosen $N_{\rm obs}=5$ so as to reduce the range of 
coordinates.  As a function of increasing $N$ (with 
$\ell\lesssim N_{\rm obs}^2$) the distribution rises smoothly 
from at about $N_{\rm obs}$, and flattens out several $e$-folds 
beyond that.  (If we were to include a prior of the form 
$P(N)\propto N^{-\alpha}$, $\alpha>1$, the distribution would be 
peaked at several $e$-folds above $N_{\rm obs}$.)  As a function 
of decreasing $\ell$, the distribution rises sharply at 
$\ell\sim N_{\rm obs}^2$, and then decreases slowly until the 
minimum-allowed value of $\ell$ is reached.  Note that, unlike the 
causal patch, apparent horizon cutoff, and fat geodesic measures 
studied in \cite{BFLR}, the CAH+ measure is free of divergences in 
the $(\ell,N)$ plane.

\subsection{Anthropic selection} 
\label{sec:anthropic}

We are interested in the distributions of values of $\Lambda$ and 
$N$ measured by observers like us.  The restriction to observers
like us is important because we wish to test these distributions 
against the value of $\Lambda$ and lower bound on $N$ that we 
actually observe, and such comparisons are meaningful only insofar 
as we can assert that the values we observe should be considered as 
randomly drawn from the predicted distributions.  Insofar as our 
presence to measure these quantities correlates with their physical 
values, these correlations must be taken into account.     

What is meant by the qualification ``like us'' defines the precise 
hypothesis that is being tested, and is ultimately constrained by 
our understanding and acumen.  According to the ``principle of 
mediocrity'' \cite{AV94,GV07}, we should consider ourselves typical 
in (that is, randomly drawn from) any class of observers to 
which we belong, unless we have evidence to the contrary.  In the 
latter case, the class should be correspondingly narrowed.  As more 
data is accumulated and our models are improved, we will be able to 
specify a narrower reference class of observers and make more 
accurate predictions.

Here we simply assume that the density of observers like us at a 
given FRW proper time $\tau$ is proportional to the number of 
galaxies that passed a certain mass threshold $M$ a certain proper 
time $\Delta\tau$ before $\tau$.  (We do not distinguish between
non-relativistic baryons and cold dark matter, and by ``galaxy'' 
we refer to both the visible structure and the halo.)  The basic 
idea is that a galaxy should have a certain minimum mass, to 
permit efficient star formation and likewise to have produced and 
retained heavy elements, and should have existed for a certain 
time, to permit planetary and biological evolution.  With respect
to bubbles with negative cosmological constant, we pursue two 
possibilities:
\begin{itemize}
\item[($i$)] in this case we do not include FRW times after 
$\tau_{\rm turn}$, the time of scale-factor turnaround, presuming 
that the typically-high merger rate in such circumstances is hazardous 
to stable stellar systems in which observers like us can arise,
\item[($ii$)] in this case we include all times, ignoring the above 
effect.
\end{itemize}
To be concrete, we take $M=10^{12}$ solar masses and 
$\Delta\tau = 5$ Gyr.  (These anthropic assumptions are the same 
as those adopted and further explained in \cite{DSGSV}.)

To determine the rate at which galaxies exceed the mass $M$, 
we work in the Press-Schechter formalism \cite{PS}.  The main 
result of this approach is the collapse fraction (of 
non-relativistic matter) into galaxies of mass greater than or 
equal to $M$ by proper time $\tau$,
\beq
F_{\rm c}(M,\,\tau) = {\rm erfc}\!\left[
\frac{\delta_{\rm c}(\tau)}{\sqrt{2}\,\sigma_{\rm rms}(M,\tau)}
\right] ,
\label{Fc}
\eeq 
where erfc denotes the complimentary error function, 
$\sigma_{\rm rms}(M,\tau)$ is the amplitude of a root-mean-square 
(rms) density perturbation on a comoving scale enclosing a 
mass $M$ (according to the linearized equation of motion), and 
$\delta_{\rm c}(\tau)$ corresponds to the amplitude that a density 
perturbation must have reached, according to the linearized 
equation of motion, for it to correspond to a collapsed spherical 
top-hat over-density at time $\tau$ (the spherical top hat being 
evolved according to a non-linear analysis).  The collapse 
threshold $\delta_{\rm c}$ is in general a function of time, 
however for our purposes it is sufficient to use the 
late-time asymptotic values   
$\delta_{\rm c} = 1.72$ for $\Lambda < 0$, 
$\delta_{\rm c} = 1.50$ for $\Lambda = 0$,  and 
$\delta_{\rm c} = 1.63$ for $\Lambda > 0$.  
(At least in the case where there is no period of late-time 
spatial-curvature domination, this approximation was found to be 
accurate at the percent level by the authors of \cite{DSGSV}).

The rate at which galaxies exceed the mass $M$ is simply the time 
derivative of $F_{\rm c}$.  Incorporating the time delay mentioned 
above, we have for the density of observers
\beq
\rho_{\rm obs}(\tau)\propto \frac{\a^3(\tau_\star)}{\a^3(\tau)}
\frac{dF_{\rm c}}{d\tau} \bigg|_{M,\,\tau-\Delta\tau} \,,
\eeq      
where it is understood that $\rho_{\rm obs}=0$ for 
$\tau>\tau_{\rm turn}$ when considering case ($i$) above.
Inserting into (\ref{Ni}), we obtain
\beq
{\cal N}_{\rm obs}(\Lambda,\,N) \propto 
P(\Lambda,\,N)\int_{\Delta\tau}^{\tau_{\rm max}}
d\tau\, e^{3N-3f[\eta(\tau)]} \,
\frac{dF_{\rm c}}{d\tau} \bigg|_{M,\,\tau-\Delta\tau}\,,
\label{Nobs}
\eeq    
where for $\Lambda<0$, we use $\tau_{\rm max}=\tau_{\rm turn}$ 
for case ($i$) and $\tau_{\rm max}=\tau_{\rm crunch}=2\tau_{\rm turn}$ 
for case ($ii$), while for $\Lambda\geq 0$, we use
$\tau_{\rm max}\to\infty$.  

Since anthropic selection will strongly suppress all values of 
$\Lambda$ except those within a very small (compared to the Planck or 
electroweak scales) window about zero, the prior distribution of 
$\Lambda$ is expected to be flat \cite{Weinberg1}.  (The conditions of 
validity of this heuristic argument have been studied in several simple 
landscape models \cite{S-PV06,Delia2,Delia3}, with the conclusion 
that it does in fact apply to a wide variety of scenarios.)  We shall 
assume that the argument is valid, so the prior probability is independent 
of $\Lambda$.  The appropriate prior distribution for the number of 
$e$-folds $N$ is less clear; here we simply lift the distribution 
obtained by \cite{FKMS}, based on randomly scanning the parameters of a 
linearized scalar-field potential.  Together these give the distribution
\beq
P(\Lambda,\,N) \propto N^{-4} \,.
\label{prior}
\eeq 

The distribution ${\cal N}_{\rm obs}$ can be evaluated numerically, 
given $\a(\tau)$, $\eta(\tau)$, and $\sigma_{\rm rms}(M,\,\tau)$.  Each 
of these quantities is approximated in Appendix \ref{sec:bigbang}, and 
during the relevant times can be expressed entirely in terms of $\HL$, 
$\tc$, and the density contrast $\sigma_{\rm ref}(M)$ evaluated at some reference time $\tau_{\rm ref}$.  It is convenient to write 
\beq
\lambda = {\rm sign}(\Lambda)\,(\HL/\tilde{H}_\Lambda)^2 \,,
\qquad {\rm and} \qquad
\tc = \tilde{H}_\Lambda^{-1} e^{3(N-\Nc)} \,, 
\eeq
where $\tilde{H}_\Lambda$ is the value of $\HL$ measured in 
our universe, $\tilde{H}_\Lambda^{-1} = 16.3$ 
Gyr,\footnote{\label{foot1} Here and throughout we use 
``WMAP+BAO+$H_0$'' maximum-likelihood cosmological parameters 
from the WMAP-7 analysis \cite{WMAP}.  In particular, we choose  
$h=0.704$, $\Omega_\Lambda = 0.728$, $\Omega_c = 0.271$, 
$\Omega_b = 0.0455$, $\Omega_k\leq 8.3\times 10^{-3}$ 
(95\% confidence level upper bound), $\tau_0=13.8$ Gyr, 
$T_{\rm CMB} = 2.725$ K, and 
$\Delta^2_{\cal R}(k=0.002\,\mbox{Mpc}^{-1}) = 2.42\times 10^{-9}$
in a $\Lambda$CDM cosmology (with three generations of massless
neutrinos).} 
and
\beq
\Nc = \frac{1}{3}\ln\!\left(3\cdot 2^{5/2}\Hd^{3/2}\tau_{\rm eq}^{1/2}
\tilde{H}_\Lambda^{-1}\right) .
\eeq
Since $\Nc$ depends on $\Hd$, its precise value in our universe 
is not known (in our simple model of inflation and reheating 
$\Hd\approx 10^3\mbox{--}10^{15}$ GeV would correspond to 
$\Nc\approx 50\mbox{--}64$).  Nevertheless, after the onset of 
non-relativistic matter domination the evolution of the universe 
depends only on the difference $N-\Nc$, and so dependence on $\Nc$ 
by itself enters only via the relative effect of the prior 
distribution $P(\Lambda,\,N)$, and this effect is approximately 
independent of $\Nc$ when $\Nc\gg 1$. For concreteness we choose 
$\Nc=60$.  Finally, we note that the data of footnote 
\ref{foot1} correspond to setting $\sigma_{\rm ref}(M)$ so that 
deep into the epoch of non-relativistic matter domination, we have
$\sigma_{\rm rms}(M=10^{12}M_\odot)=2.50\,(\tilde{H}_\Lambda\tau)^{2/3}$.

\begin{figure}[t!]
\begin{center}
\begin{tabular}{cc}
\includegraphics[width=0.44\textwidth]{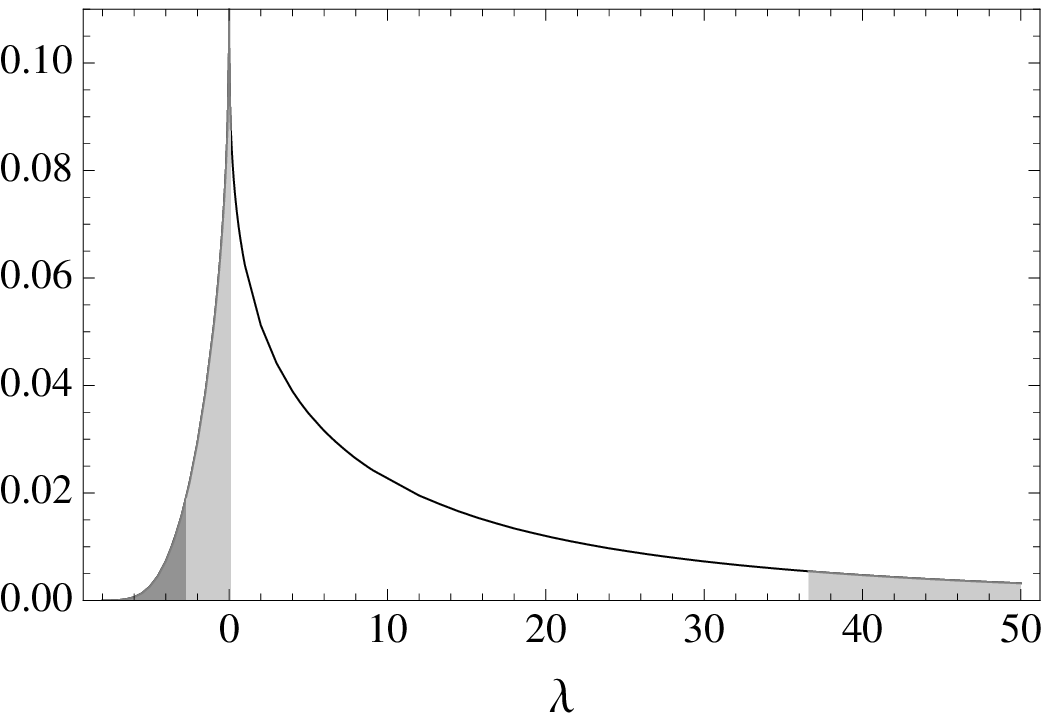} &
\includegraphics[width=0.44\textwidth]{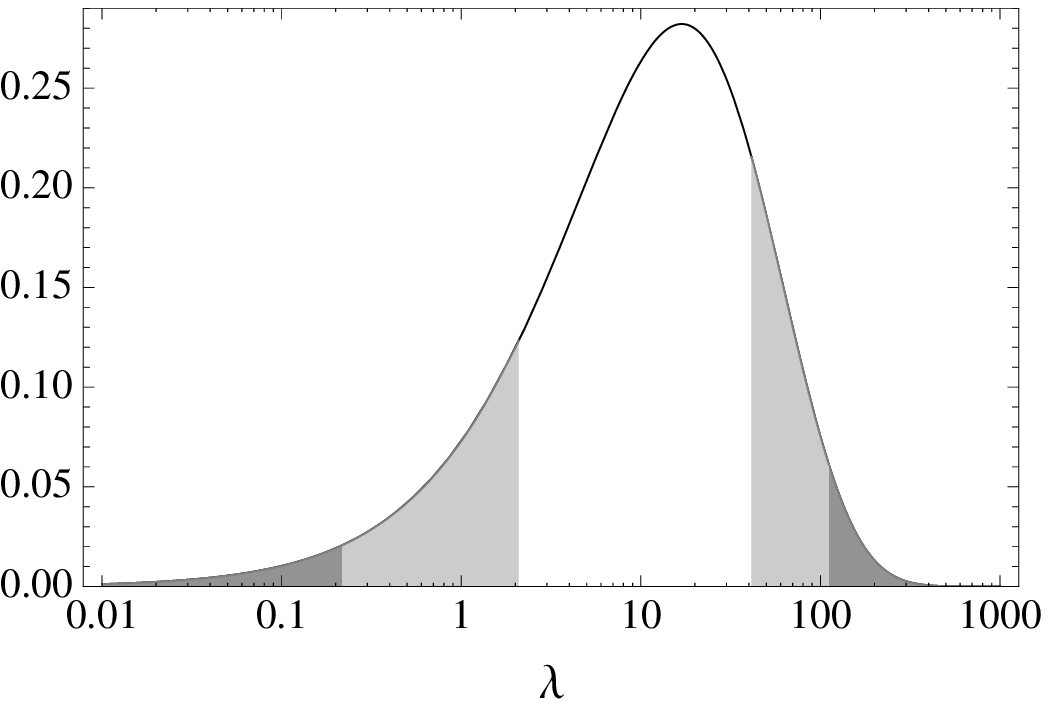} \\
\includegraphics[width=0.44\textwidth]{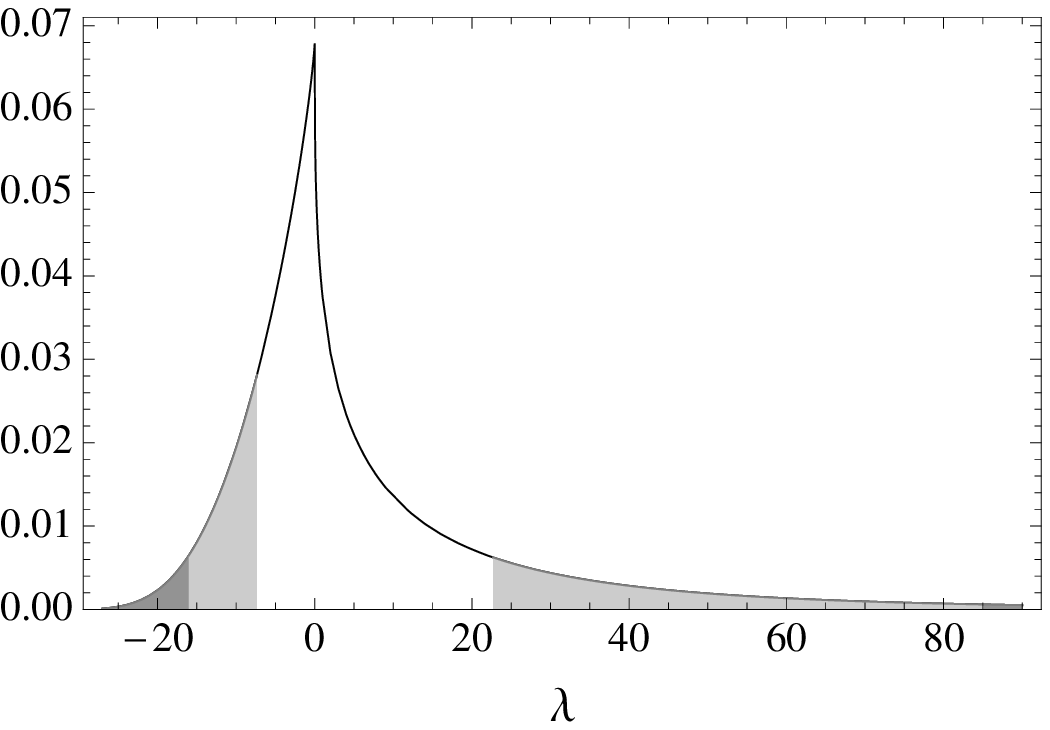} &
\includegraphics[width=0.44\textwidth]{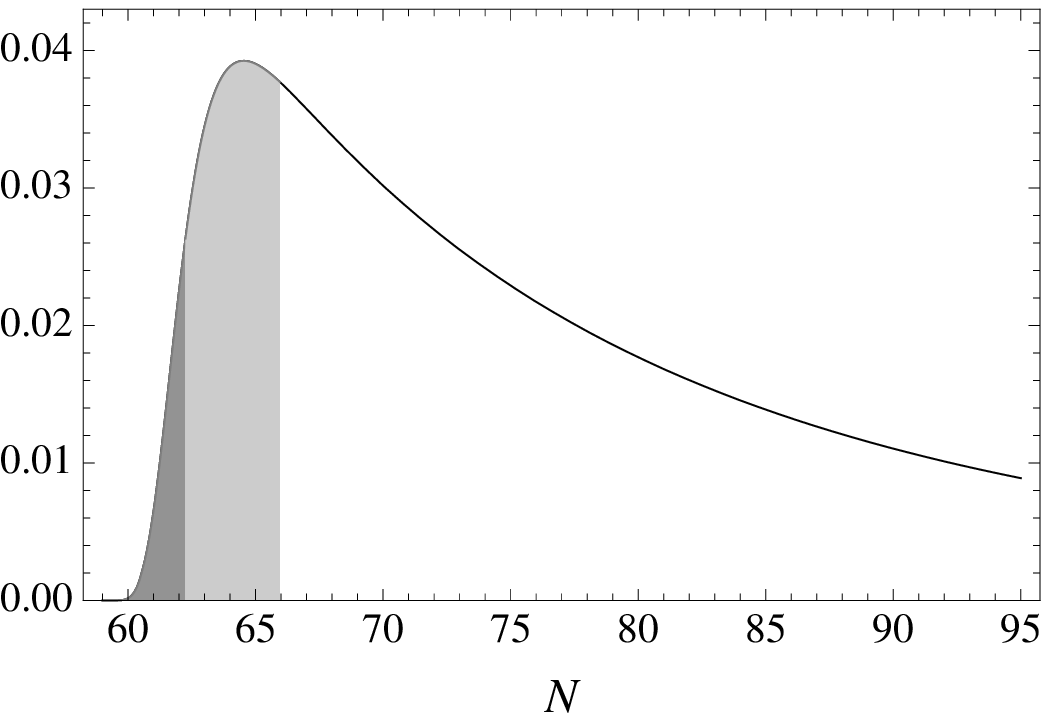}
\end{tabular}
\caption{\label{fig:L} Slices of the (normalized) distribution 
${\cal N}_{\rm obs}(\lambda,\,N)$.  The left panels display the
distribution over all $\lambda$, for anthropic case ($i$) (top) 
and case ($ii$) (bottom), for $N=70$.  The right panels display 
the distribution over positive $\lambda$ (top), again using $N=70$, 
and the distribution over $N$ (bottom), for $\lambda=1$.  In each
case we use $\Nc=60$.  Observation gives $\lambda=1$ and 
$N\geq 62.1$.  Regions more than one and two standard deviations 
away from the median indicated by shading.}
\end{center}
\end{figure}

In Figure \ref{fig:L} we plot slices of the distribution
${\cal N}_{\rm obs}(\lambda,N)$.  Recall that the parameter 
$\lambda$ has been defined so that our universe corresponds to 
$\lambda=1$.  Meanwhile, the observational constraint on the 
curvature parameter $\Omega_k = (\a H)^{-2}$ implies a constraint 
on $\Delta N$; in particular $\Delta N \geq 2.1$ (95\% confidence 
level lower bound), which for $\Nc=60$ corresponds to $N\geq 62.2$. 
(Here we have combined the data of footnote \ref{foot1} with the 
scale-factor approximations of Appendix \ref{sec:bigbang}, with 
the present FRW time corresponding to just after the onset of 
cosmological-constant domination).  We see that the distributions 
provide a good fit to the observed value of $\lambda$, whether we 
allow for negative $\lambda$ or not (indeed, regardless of 
whether we impose anthropic condition ($i$) or not), at least when 
there is much more than sufficient inflation, as in our universe. 
Note that the sharp peak at $\lambda=0$ is not a divergence.  The 
results are consistent with what is found in Section \ref{sec:geometry}, 
with the distribution being more sharply peaked about 
$\Lambda=0$ than in that analysis because we have conditioned on 
the formation of galaxies like ours, which becomes inhibited for 
large positive values of $\lambda$.

Regarding the distribution of $N$, we see the probability is 
highly suppressed for $N<\Nc$, in accordance with what was found in 
Section \ref{sec:geometry}.  As in that analysis, the distribution 
would rise and asymptote to a constant value as $N$ is increased 
over the next several $e$-folds; however we have included the prior 
(\ref{prior}), which selects for smaller values of $N$, hence the 
peaked distribution of Figure \ref{fig:L}.  Given the 
location of the peak and its long, large-$N$ tail, it is evident 
that the distribution is consistent with the present observational 
limit, $N\geq 62.2$ for $\Nc=60$ (corresponding to 
$\Omega_k \leq 8.3\times 10^{-3}$, see footnote \ref{foot1}).  Our
ability to detect spatial curvature (for instance, via its effects 
on the CMB) is cosmic-variance limited \cite{var1,var2}, and it is 
worthwhile to ask the following question.  Given this model, and 
given the present observational limit, what is the likelihood that 
$N$ lies in the range amenable to future detection?  To quickly 
estimate the answer, following \cite{FKMS} we crop the distribution 
at the present observational 
bound, and compute the fraction of the remaining distribution for
which $\Omega_k \leq 10^{-4}$ ($N\geq 64.3$), finding that it is 
0.077.  Thus, there is reasonable hope for a future detection.  These 
results are similar to those found in \cite{DSS} for the distribution
of $N$ in the scale-factor cutoff measure, and the discussion there
concerning other possibilities for the prior distribution $P(N)$ 
applies here as well.  
  
\begin{figure}[t!]
\begin{center}
\includegraphics[width=0.5\textwidth]{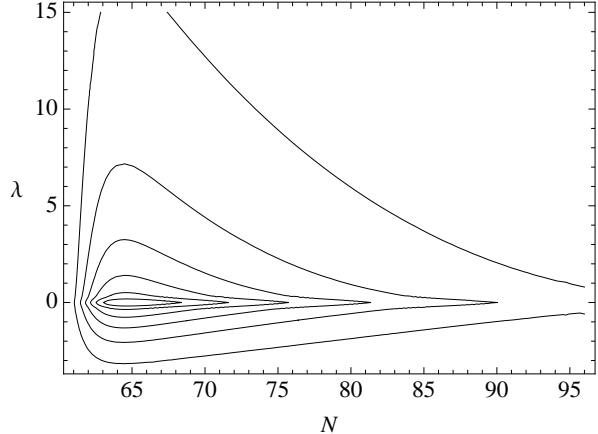}
\caption{\label{fig:NL} Contour plot of the distribution 
${\cal N}_{\rm obs}(\lambda,N)$, for case ($i$), assuming 
$N_\Lambda=60$.}
\end{center}
\end{figure}

In Figure \ref{fig:NL} we provide a contour plot of 
${\cal N}_{\rm obs}(\lambda,N)$, in the $(N,\lambda)$ plane, for
case ($i$).  Although it might appear as if constant-$N$ slices 
might represent different distributions of $\lambda$ as one surveys 
increasing $N$, in fact (for $N$ significantly larger than $\Nc$) all 
that changes is the overall normalization, according to the 
$N$-dependence of the prior (\ref{prior}).

\section{Summary of measure proposals}
\label{sec:review}

To provide context for the results of Section \ref{sec:phenomenology},
we briefly summarize the major qualitative, phenomenological 
characteristics of several measure proposals.  In many cases we 
simply restate the conclusions of \cite{BFLR} (however we draw 
different conclusions about the scale-factor cutoff measure), and, as 
in that analysis, we focus on the predictions for the cosmological 
constant $\ell$ and the number of $e$-folds $N$ relative to 
$N_{\rm obs}$ (we use the notation of Section \ref{sec:geometry}).  
We restrict attention to a subset of measures that have been 
demonstrated to avoid the youngness paradox \cite{Tegmark:2004qd,
Guth:2007ng,Bousso:2007nd}, the Boltzmann brain problem 
\cite{Dyson:2002pf,Albrecht:2004ke,Page:2006dt} (at least for 
landscapes with sufficiently rapid decay rates), and the $Q$ and 
$G$ catastrophes \cite{Feldstein:2005bm,Garriga:2005ee,
Graesser:2006ft}. 

The geometric cutoff measures that have been discussed so far can be 
divided into two broad categories: global and local measures.  Global 
measures introduce a cutoff at a fixed value of some global time 
variable $t$, be it the proper time \cite{GLL,Vilenkin:1994ua,
Vilenkin:1995yd,Linde:2006nw}, scale-factor time \cite{Vilenkin:1995yd,
Linde:2006nw,DSGSV}, lightcone time \cite{Bousso:2008hz}, 
or CAH time, and find the relative numbers of events in the limit 
$t\to\infty$.   An attractive property of these measures is that the 
resulting distributions do not depend on the choice of the comoving 
region that is being sampled, reflecting the attractor behavior of 
eternal inflation.  (They do depend on the choice of the time 
variable $t$.)  Local measures sample a spacetime region in the 
vicinity of a given worldline.  The first proposal of this kind is the 
causal patch measure \cite{Bousso:2006ev}, which counts events that 
occur within the causal patch of the wordline.  (The causal patch 
could either be the past lightcone of the future endpoint of the 
worldline, or the causal diamond of the worldline; the two choices 
give essentially the same phenomenology).  This measure was motivated 
by the idea of ``horizon complementarity'' \cite{Dyson:2002pf,Bousso:2006ge}, 
suggesting that semi-classical gravity can be trusted only within a 
causal patch of spacetime accessible to a single observer.  Other 
local measures include the apparent horizon \cite{BFLR} and fat 
geodesic \cite{Bousso:2008hz} measures, which sample respectively the 
region within the apparent horizon and within a fixed geometric 
distance of the worldline.

All local measures are sensitive to the choice of the initial vacuum 
where the geodesic begins, so one needs to consider an ensemble of 
observers with different initial conditions.  Without specifying
such an ensemble, these measures remain essentially undefined.

The key to specifying the appropriate ensemble may be provided by the 
recently-discovered duality \cite{GL-Duality} between local and 
global measures: the local measure is equivalent to the global measure 
if the ensemble of initial vacua for the local measure is given by the 
attractor distribution of the corresponding global measure.  (This 
global-local duality is somewhat limited.  As we discuss below,
its validity with respect to the fat geodesic measure and scale factor
cutoff measure breaks down in AdS vacua.)  Thus, one can 
define the causal patch, apparent horizon, and fat geodesic measures by 
requiring that the initial distribution should be taken, respectively, 
from the global lightcone, CAH, and scale factor cutoff measures.  
Note however that these measures rely on a global picture of 
spacetime, so that these definitions seem to undermine the 
initial motivation for the causal patch measure.

Some alternative proposals for specifying the initial distribution can 
be found in \cite{Vitaly}.  This distribution could also be determined 
by the wave function of the universe \cite{Bousso:2006ev}.
Whatever prescription is chosen, one can expect that the arguments we 
gave concerning the ``prior'' distribution of $\Lambda$ and $N$ should 
apply here as well, allowing to predict distributions for these 
quantities regardless of the precise nature of the ensemble.  In the 
rest of this section we discuss some qualitative features of the 
resulting distributions, focusing on the divergent, or "runaway", 
behavior exhibited by some of the measures.

We begin with the (local) apparent horizon cutoff measure.  According 
to Figure 4 of \cite{BFLR}, when $\ell$ is positive this measure
predicts a distribution ${\cal N}_{\rm obs}$ that is peaked toward 
$\ell = 0$ (with $\ln(\lambda)$ peaked at order unity) and at
$N\sim N_{\rm obs}$ (if the prior $P(N)$ features power-law preference 
for smaller $N$).  When $\ell$ is negative, ${\cal N}_{\rm obs}$ 
features a runaway toward decreasing $|\ell|$ and decreasing $N$.  
Although the results we are quoting ignore anthropic selection effects, 
it is hard to see how anthropic selection could mitigate the runaway 
toward small $|\ell|$.  Therefore only the discreteness of 
(anthropic) landscape cuts off the divergence, so that this measure 
predicts the overwhelming majority of observers to live in some vacuum 
with a negative cosmological constant very near to zero, in conflict 
with our observation.   

The causal patch measure exhibits the same qualitative behavior for 
$\ell < 0$ \cite{BFLR}.  At the same time, this measure features
a runaway toward small $\ell$, when $\ell >0$.  The runaway is much
weaker for positive $\ell$ (where it increases like $\ell^{-1}$) than 
for negative $\ell$ (where it increases like  $|\ell|^{-2}$).  This 
implies once again that the overwhelming majority of observers live 
in some negative-$\ell$ vacuum.

The authors of \cite{BFLR} turn the runaway for $\ell>0$ 
into a prediction: assuming the runaway for $\ell<0$ is
somehow resolved (and in a way that does not leave behind the 
milder, but still discouraging preference for negative $\ell$ found in 
\cite{MPS}), we should expect to observe the smallest (positive) value 
of $\ell$ among the subset of vacua that are not extraordinarily rare 
in the multiverse, and in which observers are not too strongly 
suppressed.  Then $\Lambda$ is expected to be on the order of one over 
the number of such vacua in the landscape (in Planck units).

This is an interesting attempt to relate the size of the landscape to 
the observed value of $\Lambda$.\footnote{Stronger runaway 
behavior in $\Lambda$ has been explored in \cite{LindeVanchurin}.}  
The danger here is that the distribution of $\Lambda$ is expected to be 
very ``jagged'' on the smallest energy scales, with one value of 
$\Lambda$ heavily preferred or heavily disfavored next to a 
nearest neighbor \cite{S-PV06,Delia2,Delia3}.  If the landscape is 
sufficiently enormous, one might expect this jaggedness to occur 
only on scales that are very small compared to any scale of interest, 
leaving an approximately smooth, flat distribution when averaged over an 
intermediate scale.  If however $\Lambda$ is on the order of one over 
the number of vacua in the landscape, then this jaggedness will be 
relevant to the prediction of $\ell$, perhaps giving large 
preferential weight to values that would otherwise seem hostile to 
observers.  Further investigation of this issue is required to draw more definite conclusions.  

The properties of the fat geodesic measure can be read off Figure 5 
of \cite{BFLR}.  When $\ell$ is positive this measure predicts a 
distribution ${\cal N}_{\rm obs}$ that is peaked toward $\ell = 0$ 
(with $\ln(\ell)$ peaked at order unity) and $N\sim N_{\rm obs}$ (if 
the prior $P(N)$ features power-law preference for smaller $N$).  When 
$\ell$ is negative, ${\cal N}_{\rm obs}$ features a runaway toward 
$\tau_{\rm obs}\to\tau_{\rm crunch}$, where $\tau_{\rm crunch}$
is the FRW proper time of the big crunch singularity in these cosmologies.
This runaway stems from the fact that a fixed physical volume encloses 
a diverging quantity of matter as the scale factor tends toward zero.  
It is not hard to see how anthropic selection alleviates the problem,
since the universe will become hotter and denser as the scale factor
contracts, and the environment will at some point become hostile to 
observers.  This formally cuts off the divergence in the distribution
as $\tau_{\rm obs}\to\tau_{\rm crunch}$, but to what extent the 
resulting distribution prefers negative $\ell$ depends on the
anthropic criteria that are added to the computation.

It should be noted that the apparent horizon and fat geodesic measures 
violate the causality condition mentioned in the introduction, 
which requires that the cutoff surfaces should be spacelike.  For the 
causal patch measure the cutoff surface is null and can be made 
spacelike by an infinitesimal deformation.

We finally discuss the global scale-factor (SF) cutoff measure.  It 
imposes a cutoff at a fixed value of the SF time, which was defined in 
\cite{DSGSV,DeSimone:2008if} as (the logarithm of) the volume expansion 
factor along the geodesic congruence.  As it stands, this definition is 
not satisfactory, since the resulting measure is sensitive to the 
details of structure formation in thermalized regions of spacetime 
\cite{Bousso:2008hz}.  Moreover, the cutoff hypersurface obtained with 
this measure have timelike segments in the vicinity of 
gravitationally-collapsed structures and in bubbles with negative 
vacuum energy after scale-factor turnaround.  This can be dealt with 
by augmenting the cutoff hypersurface with the future lightcones of all 
points on the cutoff, as with the ``+'' prescription of this paper.  
(This and other possibilities are discussed in \cite{DeSimone:2008if}.)  
The resulting measure can be called SF+.  

\begin{figure}[t!]
\begin{center}
\begin{tabular}{ccc}
\includegraphics[width=0.33\textwidth]{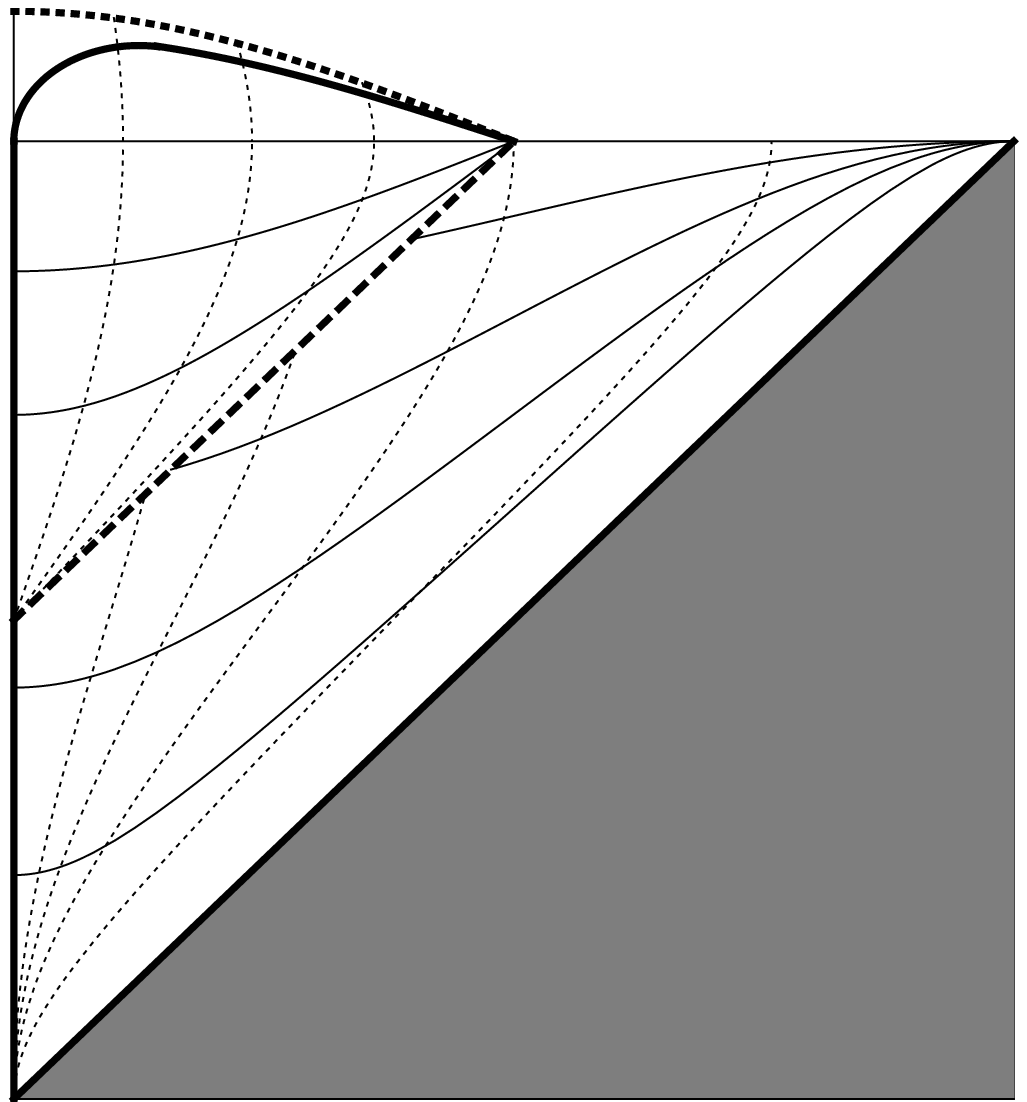} & \phantom{ssss} &
\includegraphics[width=0.33\textwidth]{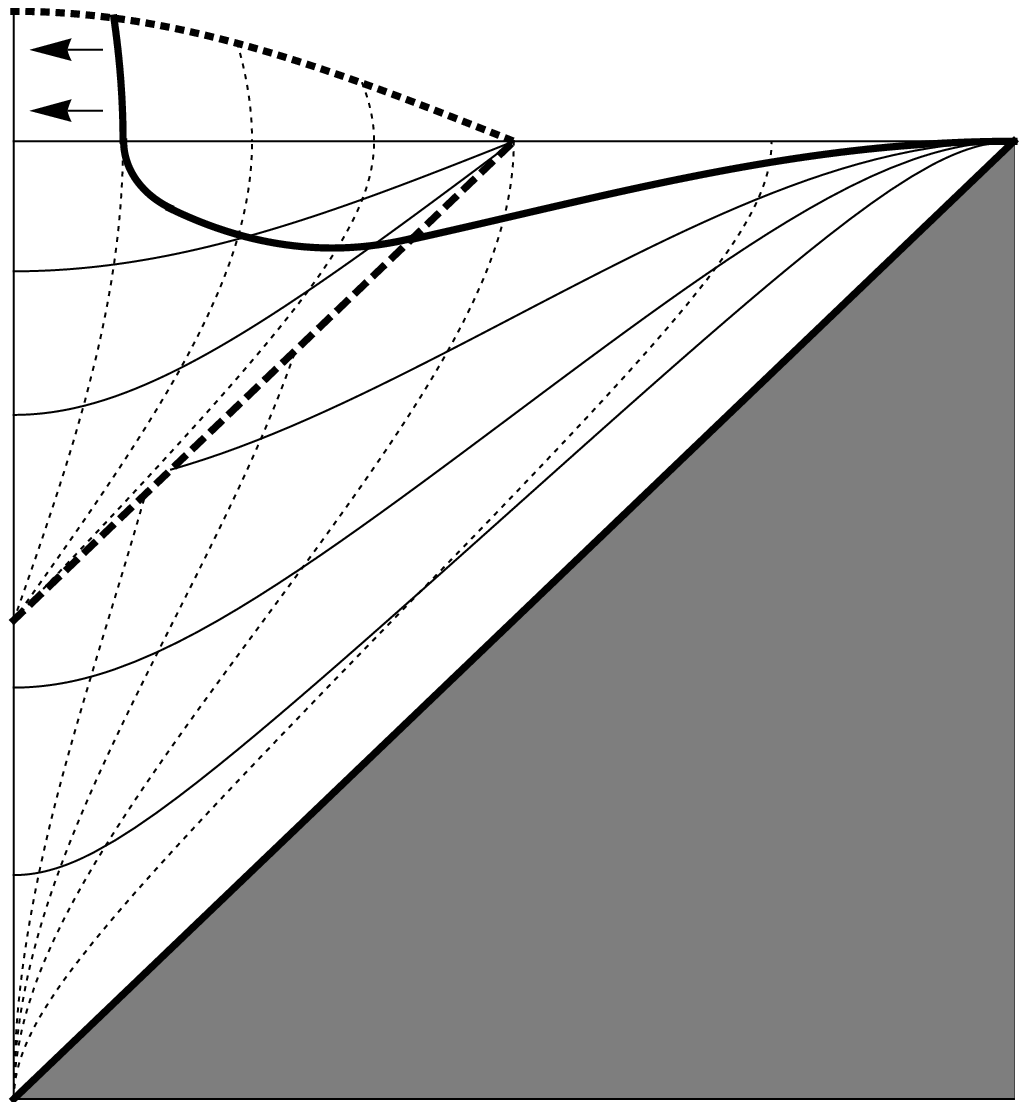}
\end{tabular}
\caption{\label{fig:comparison} Toy conformal diagrams of a crunching
bubble indicating the cutoff implied by the fat geodesic (left panel) 
and scale-factor cutoff (right panel) measures.  The symbolism is the
same as in Figure 1, but with the future boundary of the bubble given 
by a big-crunch singularity (thick, dotted line).  The fat 
geodesic focuses on an arbitrarily small region surrounding a worldline 
at the left edge of the diagram, which grows to enclose a diverging 
conformal volume as the scale factor goes to zero.  The cutoff 
implied by the scale-factor cutoff measure could enclose a smaller
volume, depending on its definition, as indicated by the arrows.}
\end{center}
\end{figure}

In bubbles with positive vacuum energy, we expect the predictions of 
the SF+ measure to be very similar to those of the fat geodesic 
measure.  Indeed, given the approximations of \cite{BFLR} (in 
particular, disregarding inhomogeneities caused by structure formation), 
they make the same predictions.  Thus Figure 5 of \cite{BFLR} also 
describes the distribution of $\ell>0$ and $N$ for the SF+ measure, 
and we see there are no runaways.  When $\ell$ is negative, the two 
measures make the same predictions during the expanding phase of bubble evolution, but differ after scale-factor turnaround.  This can be seen 
by referring to Figure \ref{fig:comparison}.  Whereas the fat geodesic 
expands to enclose a diverging quantity of matter (when represented in 
the conformal diagram), the SF+ cutoff will enclose a finite quantity 
of matter, because that is all that remains along the defining 
congruence after the cutoff has been imposed outside of the crunching 
bubble.  Thus, the SF+ measure is free of runaways in the $(\ell,N)$ 
plane.

\section{Discussion and conclusions}
\label{sec:discussion}

Our goal in this paper was to investigate the phenomenological 
properties of the CAH+ measure.  We found that this measure does not 
suffer from any known pathologies, such as the youngness paradox, $Q$ 
or $G$ catastrophes, or the Boltzmann brain problem (assuming that 
anthropic vacua in the landscape have sufficiently fast decay rates).  
The distribution for the cosmological constant $\Lambda$ derived from 
the CAH+ measure is in a good agreement with the observed value, and 
the distribution for the number of inflationary $e$-foldings $N$ 
satisfies the observational constraint.  We found also that this 
measure assigns a non-negligible probability to a detectable negative 
curvature in the present universe.  By its construction, the CAH+ 
measure satisfies the causality condition, requiring that the cutoff 
hypersurfaces should be spacelike or null.  

The properties of the CAH+ measure are similar to those of the SF+ 
measure (that is, of the scale-factor measure with cutoff surfaces 
modified as in the "+" prescription of this paper).  In fact, of all 
the measures that have been discussed so far, these are the only two 
that agree with the available data and have no pathological features.  
In contrast, the causal patch, fat geodesic, and (local) apparent 
horizon cutoff measures give strong preference to negative values of 
$\Lambda$, and thus are in conflict with observations \cite{BFLR}.  
Restricting to positive values of $\Lambda$, the causal patch measure 
gives a non-normalizable distribution, diverging towards $\Lambda=0$.  
The divergence can be cut off due to the discrete character of the 
landscape, provided that there are no anthropic vacua with $\Lambda=0$.  
The maximum of the distribution is then shifted towards values 
comparable to the scale of discreteness, and the predicted value may 
be unfavorably affected by the jaggedness of the distribution on that 
scale.  It should also be mentioned that all local measures are 
sensitive to the initial conditions and remain essentially 
undefined until the ensemble of initial states is specified.

For completeness, we mention some measures that have not been discussed 
in the main text.  The proper time \cite{GLL,Vilenkin:1994ua,Linde:2006nw}, 
pocket-based \cite{GSVW}, and stationary \cite{Linde:2008xf} measures 
all suffer from $Q$ and $G$ catastrophes.  In addition, the proper time 
measure is subject to the youngness paradox, and the pocket based 
measure to the Boltzmann brain problem.

Throughout this paper we assumed the spacetime to be $(3+1)$-dimensional.  However, in general we expect the string theory landscape to allow 
parent vacua to nucleate daughter vacua with different effective 
dimensionality \cite{Linde:1988yp,BlancoPillado:2009di,
Carroll:2009dn,BlancoPillado:2010uw}.  Predictions of the scale factor 
cutoff measure in a transdimensional landscape have been studied in 
\cite{SPV10}, with the conclusion that in cases when the highest number 
of spatial dimensions in inflating vacua is $D_{\rm max}>5$, this 
measure strongly disfavors large amounts of slow-roll inflation in the 
bubbles and predicts low values of the density parameter $\Omega$, in 
conflict with observations.  This problem is avoided if instead of the 
scale factor measure one uses the volume factor (VF) cutoff, where the 
cutoff surfaces are surfaces of constant volume expansion factor.  
(These surfaces are the same as the constant scale factor surfaces in 
3+1 dimensions, but are generally different in a transdimensional 
multiverse.)  In order to comply with the causality condition, the 
modified VF+ measure can be defined by adding the ``+'' prescription.

The situation with the CAH+ measure is similar.  It generally 
predicts low values of $\Omega$ in the transdimensional case.  The 
analog of the VF+ measure in this case is the CNAH+ measure, in which 
the cutoff surfaces are the surfaces having a constant number of 
apparent horizons per equal comoving volume (with added ``+'' 
prescription).  It is not presently clear how this measure can be 
related to a UV cutoff in the holographic boundary theory.  This 
remains a topic for future research.

\acknowledgments

We are grateful to Ben Freivogel, Jaume Garriga and Daniel Harlow for 
useful discussions.  MPS is supported by the Stanford Institute for 
Theoretical Physics.  AV is supported by the NSF grant PHY-0855447.

\appendix

\section{Evolution of the CAH through CDL bubble nucleation}
\label{sec:CDL}

Here we compute the evolution of the CAH time $\theta$ along a 
congruence of timelike geodesics that begin in the parent vacuum 
and enter the bubble.  Disregarding bubble collisions, the spacetime 
in the vicinity of a bubble can be represented by the spatially-flat 
de Sitter chart, 
\beq
ds^2 = -dt^2 +e^{2\Hp t}\left(dr^2+r^2\,d\Omega_2^2\right) ,
\label{flatdS}
\eeq
where $\Hp$ is the parent-vacuum Hubble rate.  We consider a 
congruence $\gamma_r$ of geodesics parametrized by $r=$ 
constant in the parent vacuum.  As mentioned in the main text, the 
bubble geometry is of the form
\beq
ds^2 = -d\tau^2 + \a^2(\tau)\Big[d\xi^2
+\sinh^2(\xi)\,d\Omega_2^2\Big] \,,
\label{FRW}
\eeq
where here we focus exclusively on the scale-factor solution
\beq
\a(\tau) = \Hd^{-1}\,\sinh(\Hd\tau) \,.
\label{a}
\eeq  
We take the bubble to nucleate with a negligible initial radius at 
the parent-vacuum coordinates $(t,\,r)=(0,\,0)$, the future lightcone 
of which corresponds to the hypersurface $\tau=0$ in the bubble.  
Furthermore we work in the thin-wall approximation, with the bubble 
wall at the aforementioned future lightcone, and neglect any effect of
bubble-wall tension on the trajectories of geodesics that pass through 
it.    

To propagate the congruence $\gamma_r$ into the bubble, it will help
to embed the spacetime in a (4+1)-dimensional Minkowski space, for 
which we write the line element
\beq
ds^2=-dT^2+dS^2+dR^2+R^2\,d\Omega_2^2 \,.
\eeq
The geometry of the parent vacuum is induced on the hyperboloid 
$-T^2+S^2+R^2 = \Hp^{-2}$, which can be accomplished with the 
embedding
\bea
T &=& \Hp^{-1}\sinh(\Hp t)+\frac{1}{2}\Hp\, r^2\, e^{\Hp t} 
\label{embedp1}\\
S &=& \Hp^{-1}\cosh(\Hp t)-\frac{1}{2}\Hp\, r^2\, e^{\Hp t} \\
R &=& r \, e^{\Hp t} \,, \phantom{\frac{1}{2}} \label{embedp3}
\eea  
where we have suppressed coordinates on the unit two-sphere, which
do not concern us here.  A general spacetime of the form (\ref{FRW}) 
can also be embedded in Minkowski space, but for concreteness we 
focus on the specific, early-time solution (\ref{a}).  This can be 
embedded via 
\bea
T &=& \Hd^{-1}\,\sinh(\Hd\tau)\cosh(\xi) \\
S &=& \Hd^{-1}\,\cosh(\Hd\tau) +\Hp^{-1}-\Hd^{-1}\\
R &=& \Hd^{-1}\,\sinh(\Hd\tau)\sinh(\xi)\,.
\eea  
The bubble geometry then corresponds to that induced on the 
hyperboloid 
\beq
-T^2+\big(S+\Hd^{-1}-\Hp^{-1}\big)^2+R^2=\Hd^{-2}\,, 
\eeq
which has been shifted from the origin so that it intersects the 
parent-vacuum hyperboloid at $S=\Hp^{-1}$, where we place the bubble 
wall.  For the purpose of matching the two coordinate systems it is 
convenient to also introduce a spatially-flat de Sitter chart in the 
bubble.  The line element is of the form (\ref{flatdS}), but with 
$\Hp\to\Hd$, and likewise the embedding is of the form 
(\ref{embedp1})--(\ref{embedp3}), but with $\Hp\to\Hd$ followed by 
$S\to S+\Hp^{-1}-\Hd^{-1}$.  

\begin{figure}[t!]
\begin{center}
\includegraphics[width=0.4\textwidth]{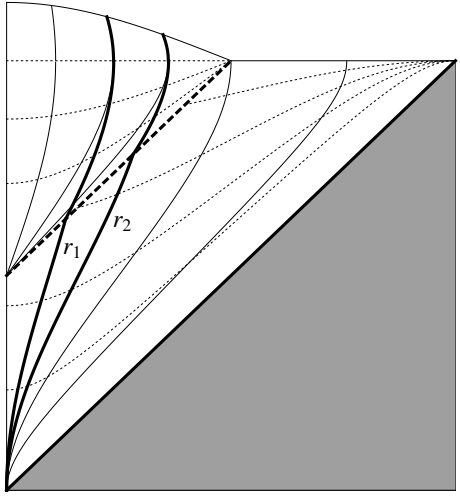}
\caption{\label{fig:geodesics} Two initially-comoving geodesics, $r_1$ 
and $r_2$ (dark, solid curves), propagating from the parent vacuum into 
the bubble.  The bubble wall corresponds to the thick, dashed line; 
dotted lines correspond to constant-$t$ (constant-$\tau)$ surfaces, 
while solid lines correspond to constant-$r$ (constant-$\xi$) surfaces, 
in the parent vacuum (bubble).}
\end{center}
\end{figure}

Our calculation of the propagation of the congruence $\gamma_r$ into 
the bubble follows the analysis of \cite{VW97} (for more background see 
\cite{matching}).  Figure \ref{fig:geodesics} illustrates the dynamics, 
projected onto a conformal diagram of CDL bubble nucleation (the 
spatially-flat de Sitter chart of the parent vacuum covers only the 
upper-left half of the diagram).  Initially-comoving geodesics (with 
respect to the spatially-flat parent-vacuum frame) encounter the bubble 
wall, are boosted with respect to the bubble coordinates, but 
nevertheless rapidly asymptote to comoving in the bubble FRW frame, due 
to redshifting during inflation in the bubble.

In terms of the two spatially-flat charts, the trajectory of the bubble 
wall, specified by the intersection of the hypersurface $S=\Hp^{-1}$ with 
the two hyperboloids, is given by
\bea
\Hp r_{\rm w} &=& 1-e^{-\Hp t_{\rm w}} \,, \qquad
\Hp t_{\rm w} = \ln(\Hp\,\lambda+1) \,, \\
\Hd \r_{\rm w} &=& 1-e^{-\Hd\t_{\rm w}} \,, \qquad
\Hd \t_{\rm w} = \ln(\Hd\,\lambda+1) \,, \label{dwall}
\eea
where we use overlines to denote coordinates in the spatially-flat
chart within the bubble.  The choice of parameter $\lambda$ is 
such that either trajectory maps to the same embedding coordinate $T$
(and, by definition, $S$) for a given value of $\lambda$. 

Meanwhile, in terms of the spatially-flat slicing in the bubble, a 
general timelike geodesic (orthogonal to the unit two-sphere) takes 
the form
\bea
\Hd \r &=& \Hd \r_{\rm in} + \Big(v^{-2}+e^{-2\Hd\t_{\rm in}}\Big)^{\!1/2}
-\,\Big(v^{-2}+e^{-2\Hd\t}\Big)^{\!1/2} \label{bubblegeo1}\\
\Hd \t &=& \ln\!\left[v\sinh(\Hd s)\right] ,
\label{bubblegeo2}
\eea
where $(\r_{\rm in},\,\t_{\rm in})$ is the point along the bubble 
wall through which the geodesic passes, with 
$\Hd \r_{\rm in} = 1-e^{-\Hd\t_{\rm in}}$ in accordance with
(\ref{dwall}), $v$ is an additional integration constant, and $s$ is
the proper time along the geodesic.  The coordinate $\r_{\rm in}$ of
a geodesic just inside the bubble can be related to an element $r$ 
of the comoving congruence in the parent vacuum by matching the 
corresponding Minkowski embedding coordinates at the bubble wall.  
This gives
\beq
\r_{\rm in} = \frac{r}{1-\Hp r+\Hd r} \,.
\label{rin}
\eeq
The integration constant $v$ can be determined by equating 
the inner products between the geodesic and the normal to the 
bubble wall worldsheet, on either side of the wall.  This gives  
\beq
v = \frac{r\,(\Hp-\Hd)(2-\Hp r +\Hd r)}{2\,(1-\Hp r)} \,.
\label{v}
\eeq

We are now prepared to express the propagation of the congruence 
$\gamma_r$ into the bubble, in terms of the open--de Sitter chart
(\ref{FRW})--(\ref{a}) respecting the FRW symmetry of the bubble.  
The spatially-flat and open de Sitter charts can be related by 
identifying points on the hyperboloid in the Minkowski embedding
space; this gives
\bea
e^{\Hd\t} &=& \cosh(\Hd\tau)+\sinh(\Hd\tau)\cosh(\xi)
\phantom{\frac{1}{2}} \\
\Hd\r &=& \frac{\sinh(\Hd\tau)\sinh(\xi)}
{\cosh(\Hd\tau)+\sinh(\Hd\tau)\cosh(\xi)} \,.
\eea
We see that at sufficiently late FRW times $\tau$, the 
spatially-flat coordinates become comoving with respect to the 
open coordinates (that is, $d\r/d\tau\to 0$), and we have
\beq
e^{\Hd\t} \,\to\, e^{\Hd\tau}
\left[\frac{1}{2}+\frac{1}{2}\cosh(\xi)\right] ,\qquad
\Hd\r \,\to\, \frac{\sinh(\xi)}{1+\cosh(\xi)} \,.
\eeq
Since $\cosh(\xi)\geq 1$, the inequality $\Hd\tau\gg 1$ implies the
inequality $\Hd\t\gg 1$.  Meanwhile, for times $\Hd\t\gg 1$, the 
geodesic given by (\ref{bubblegeo1}) and (\ref{bubblegeo2}) 
asymptotes to evolution along a fixed value of the coordinate $\r$.  
Substituting (\ref{rin}) and (\ref{v}) into (\ref{bubblegeo1}), we 
find
\beq
\Hd \r \,\to\, \frac{\Hp r+\Hd r}{2-\Hp r+\Hd r} \,.
\eeq
Expressing this in terms of the bubble coordinate $\xi$, we obtain
\beq
\xi \,\to\, \ln\!\left(\frac{1+\Hd r}{1-\Hp r}\right) \,.
\label{r2xi}
\eeq

With the dynamics of the geodesic congruence $\gamma_r$ in hand, we 
can now evolve the CAH time from deep within the parent vacuum to deep 
within the bubble.  We first pick an ``initial'' constant-AH 
hypersurface $\Sigma_0$, which we take to coincide with the hypersurface 
$t=0$ in the parent vacuum.  The scale factor is constant on 
this hypersurface, so it is also a surface of constant CAH, with 
$r_{\rm CAH} = r_{\rm AH} = \Hp^{-1}$.  According to (\ref{r2xi}), 
a small, radial comoving coordinate separation $\Delta r$ 
in the spatially-flat de Sitter slicing of the parent vacuum 
translates into a comoving coordinate separation
\beq
\Delta \xi = \frac{(\Hp+\Hd)\,\Delta r}{(1-\Hp r)(1+\Hd r)} 
= \frac{(\Hp+\Hd e^{-\xi})^2\,e^\xi\,\Delta r}{\Hp+\Hd} \,,
\eeq    
at times $\Hd\tau\gg 1$ in the bubble.  To obtain the CAH in
the bubble, we start with the AH radius $r_{\rm AH}=\Hd^{-1}$, 
divide by the bubble scale factor to convert to comoving 
coordinates, and then divide by 
$\left[\sinh^2(\xi)\,\Delta\xi / r^2\Delta r\right]^{1/3}$
to convert comoving bubble coordinates to comoving parent-vacuum 
coordinates on the ``initial'' hypersurface $\Sigma_0$.  (Here we
exploit the symmetries of an ``annulus'' centered at $r=\xi=0$
to compute the comoving volume expansion factor implied by the
relation (\ref{r2xi}).)  This gives the CAH time 
\beq
\theta\equiv \frac{1}{r_{\rm CAH}} = \frac{\a(\tau)}{r_{\rm AH}}
\left[\frac{\sinh^2(\xi)}{r^2(\xi)}\frac{\Delta\xi}{\Delta r}\right]^{1/3}
\!=\,\left[\frac{(1+e^{-\xi})^2(\Hp+\Hd e^{-\xi})^4}
{32\,(\Hp+\Hd)}\right]^{1/3} e^{\xi+\Hd\tau} \,.
\label{eta1}
\eeq

Note that the CAH time depends on $\xi$, rapidly increasing for 
large $\xi$, which is a necessary condition for the CAH cutoff to 
regulate the spacetime volume of the bubble, which is divergent on 
constant-$\tau$ hypersurfaces.  Although $\theta$ is not
constant (with respect to varying $\xi$) on surfaces of constant
FRW proper time in the bubble, at late times this variation is small on 
any fixed distance scale.  In particular, the AH scale $\Hd$ is covered 
by the comoving distance 
$\Delta\xi_{\rm AH}(\tau)= \Hd^{-1}\a^{-1}(\tau)$, and so
\beq
\frac{1}{\theta}\frac{d\theta}{d\xi}\Delta\xi_{\rm AH} =
\frac{6\Hp+2(\Hp-\Hd)e^{-\xi}-6\Hd e^{-2\xi}}
{3\,(1+e^{-\xi})(\Hp+\Hd e^{-\xi})}\,e^{-\Hd\tau} \,,
\eeq
which is very small when $\Hd\tau\gg 1$.  This means that it was
unnecessary to take the $t=0$ hypersurface as the ``initial''
constant-CAH surface with respect to which we define CAH time, as
was done above.  Since CDL bubbles in de Sitter vacua never expand 
beyond the (locally-defined) comoving AH, within the vicinity of 
any such bubble the spatially-flat de Sitter slicing serves as a 
constant CAH time foliation.

Equation (\ref{eta1}) gives the result we are looking for:  the CAH 
time $\theta\equiv r_{\rm CAH}^{-1}$ (relative to the CAH time on the 
hypersurface $t=0$ in the parent vacuum) at a given point in the 
bubble, deep in the inflationary epoch.  The evolution of $\theta$ 
during a standard big bang cosmology after inflation is discussed in 
the main text.  Note that if the vacuum energy in the parent vacuum 
is significantly larger than the inflationary energy density in the 
bubble, $\Hp\gg\Hd$, we obtain
\beq
\theta=\frac{1}{2}\Hp
\left(\frac{1}{2}+\frac{1}{2}e^{-\xi}\right)^{\!2/3}e^{\xi+\Hd\tau} \,,
\label{eta2}
\eeq 
which is qualitatively accurate even for $\Hp$ rather near $\Hd$.
The prefactor $\Hp$ can be understood to represent the CAH time on 
the initial hypersurface $t=0$ in the parent vacuum; had we chosen 
this hypersurface to reside at some earlier time $t=-t_{\rm n}$, 
this prefactor would have been $\Hp\,e^{\Hp t_{\rm n}}$.  
Meanwhile, the factor in parentheses is equal to one when $\xi=0$, 
and asymptotes to $2^{-2/3}\approx 0.63$ at large $\xi$.  We  
therefore simply ignore this factor, to simplify the algebra without 
significantly changing the CAH time in the bubble.       
Accordingly, in the main text we write
\beq
\theta=\theta_0\,e^{\Hp t_{\rm n}+\xi+\Hd\tau} \,,
\label{eta3}
\eeq
where $2\theta_0$ is the CAH time on the initial hypersurface 
$\Sigma_0$, taken to reside a proper time $t_{\rm n}$ before the
point of bubble nucleation, along a comoving geodesic in the 
parent vacuum.

\section{Big bang cosmology}
\label{sec:bigbang}

We here describe the scale-factor and growth-factor evolution in 
bubbles of interest.  We assume an initial period of curvature 
domination, in accordance with matching CDL boundary conditions at 
FRW proper time $\tau=0$ (conformal time $\eta\to-\infty$), followed 
by inflation (approximated as constant vacuum-energy domination), 
instantaneous reheating yielding radiation domination, 
non-relativistic matter domination, spatial-curvature domination 
(again), and/or cosmological-constant domination.  With respect to
scale-factor evolution, in (almost) every case we approximate the 
transition between these periods as instantaneous, solving the 
Einstein field equation $H^2=(8\pi G/3)\,\rho$ and the Friedmann 
equation $\dot{\rho}=-3H(\rho+p)$ for a perfect fluid with the 
appropriate equation of state $w=p/\rho$ (for energy density $\rho$ 
and pressure density $p$), and setting the integration constants 
so as to make the scale factor $\a$ and its first time derivative 
continuous across the transition.  Approximating the growth-factor 
evolution involves a bit more finesse, described below.

\subsection{Scale factor evolution}

From FRW proper time $\tau=0$ up until the end of inflation, we 
take the scale factor to be
\beq
\a(\tau) = \Hd^{-1}\sinh(\Hd\tau) \,\to\, \frac{1}{2\Hd}e^{\Hd\tau} \,,
\eeq
where here and below the arrows indicate the late time limits, 
which in each case we take to be accurate approximations for 
matching the scale factor onto the next phase of its
evolution.  As usual, conformal time is defined 
$\int d\tau/\a(\tau)$; so that at late times we have 
\beq
\a(\eta) \,\to\, -\frac{1}{\Hd\eta} \,, \qquad 
\eta \,\to\, -2e^{-\Hd\tau} \,.
\eeq 
We assume instantaneous reheating, and thus match directly onto the
scale-factor evolution during radiation domination ($w=1/3$).  This 
gives
\beq
\a(\tau) = e^N\left(\frac{\tau}{2\Hd}-\frac{2N-1}{4\Hd^2}\right)^{\!1/2} 
\to\, e^N\left(\frac{\tau}{2\Hd}\right)^{\!1/2} \,.
\eeq
In terms of conformal time, the late-time limit corresponds to
\beq
\a(\eta) \,\to\, \frac{1}{4\Hd}\,e^{2N}\eta \,, \qquad
\eta \,\to\, e^{-N}(8\Hd\tau)^{1/2} \,.
\eeq

The above period of radiation domination gives way to a period of
non-relativistic matter domination ($w=0$), and we denote the time
of the transition $\tau_{\rm eq}$.  The actual value of 
$\tau_{\rm eq}$ is given by microphysical parameters and will not 
by itself be important to our results.  Matching the scale-factor
evolution as described above, we find 
\beq
\a(\tau) = \frac{e^N(3\tau+\tau_{\rm eq})^{2/3}}
{2^{11/6}\,\Hd^{1/2}\tau_{\rm eq}^{1/6}} \,\to\,
\frac{3^{2/3}\,e^N\,\tau^{2/3}}
{2^{11/6}\,\Hd^{1/2}\,\tau_{\rm eq}^{1/6}} \,.
\label{amatter}
\eeq
In hindsight, we recognize that the above combination of factors
is proportional to the time of the transition to spatial-curvature 
domination (in bubbles where the number of $e$-folds of inflation is 
such that this precedes cosmological-constant domination), which we 
here note is
\beq
\tc =
\frac{e^{3N}}{3\!\cdot\! 2^{5/2}\Hd^{3/2}\tau_{\rm eq}^{1/2}} \,,
\qquad
{\rm so\,\,that} \qquad
\a(\tau) \,\to \frac{3}{2}\tc^{1/3}\tau^{2/3} \,.
\label{amatter2}
\eeq
In terms of conformal time, the late-time limit corresponds to
\beq
\a(\eta) \,\to\, \frac{3}{8}\tc \eta^2
\,, \qquad
\eta \,\to\, 2\tc^{-1/3}\tau^{1/3}\,.
\eeq

Non-relativistic matter domination persists until spatial-curvature or 
cosmological-constant domination, at time $\tc$ or $\tau_{\rm\Lambda}$.  
In the first case we treat the spatial curvature as a perfect fluid 
with energy density $\a^{-2}$ and equation of state $w=-1/3$.  Matching 
onto the previous scale-factor evolution then gives
\beq
\a(\tau) = \tau + \frac{1}{2}\tc \,\to\, \tau \,.
\label{acurv}
\eeq
The time of the transition can be obtained by solving for when the 
Hubble rate implied by (\ref{amatter2}) is equal to the Hubble rate
implied by (\ref{acurv}), which confirms the result introduced with
hindsight above.  In terms of conformal time, we have
\beq
\a(\eta) = \frac{3}{2}\tc\,e^{\eta-2} 
\,, \qquad \eta = 2+\ln\!
\left(\frac{2}{3}\frac{\tau}{\tc}+\frac{1}{3}\right) .
\label{acurveta}
\eeq

If the cosmological constant is precisely zero, the scale factor
evolves according to the above curvature-dominated results all the
way to future infinity.  Meanwhile, for positive cosmological 
constant the scale factor is given by
\beq
\a(\tau) = \HL^{-1} e^{\HL\tau+\frac{1}{2}\HL\tc-1} \,,
\eeq    
where $\HL\equiv (|\Lambda|/3)^{1/2}$ (the absolute value being 
introduced for later convenience).  Note that the transition from
curvature to cosmological-constant domination occurs at 
$\tau_\Lambda \equiv \HL^{-1}-\tc/2$.  In terms of conformal time, 
we have
\beq
\a(\eta) = \HL^{-1}\!\left[
3-\ln\!\left(\textstyle\frac{3}{2}\HL\tc\right)-\eta
\right]^{-1} , \quad 
\eta = 3-\ln\!\left(\textstyle\frac{3}{2}\HL\tc\right)
-e^{-\HL\tau-\frac{1}{2}\HL\tc+1}\,.
\eeq

When the cosmological constant is negative, it is not possible to
proceed as above and model FRW scale factor evolution as if the 
cosmological constant were the only important contribution to the 
energy density.  Therefore, for this case we include both spatial 
curvature and cosmological constant in the Einstein field equation.
The solution that matches onto the scale-factor evolution at the 
end of matter domination for generic parameters $\tc$ and $\HL$ 
involves a complicated integration constant.  Therefore we 
use the approximation 
\beq
a(\tau) = \HL^{-1}\sin\!\Big[
\HL(\tau-\tc)+\arcsin\!\left(\textstyle\frac{3}{2}\HL\tc\right)
\Big] ,
\eeq
for which the scale factor is continuous at $\tau=\tc$ for
generic parameters, and smooth when $\tc\ll\HL^{-1}$.  This
approximation for the phase of the sine function becomes poorer as 
the product $\HL\tc$ approaches 2/3 (when $\HL\tc$ is larger than 
this we use another solution for the scale factor), however the 
discrepancy becomes unimportant as $\tau$ becomes much larger than 
$\tc$.  In terms of conformal time, we have
\bea
\a(\eta) &=& \HL^{-1}\sin\!\bigg[
2\arctan\left\{e^{\eta-2}\tan\!\left[\frac{1}{2}\arcsin\!
\left(\textstyle\frac{3}{2}\HL\tc\right)\right]\right\}\bigg] \\
\eta &=& 2+\ln\!\left\{\tan\!\left[
\frac{1}{2}\HL(\tau-\tc)+\frac{1}{2}\arcsin\!
\left(\textstyle\frac{3}{2}\HL\tc\right)\right]
\cot\!\left[\frac{1}{2}\arcsin\!
\left(\textstyle\frac{3}{2}\HL\tc\right)\right] \right\} . \,\,\,\, 
\eea
After the onset of negative cosmological-constant domination, the 
scale factor eventually stops growing and subsequently decreases with 
time.  The subsequent evolution then corresponds to the time reversal 
of the evolution described above, reflected about the 
``turnaround'' time 
\beq
\tau_{\rm turn} = \tc + \frac{\pi}{2\HL} - \HL^{-1}
\arcsin\!\left(\textstyle\frac{3}{2}\HL\tc\right) \,, \quad
\eta_{\rm turn} = 2 + \ln\!\left\{
\cot\!\left[\frac{1}{2}\arcsin\!
\left(\textstyle\frac{3}{2}\HL\tc\right)\right] \right\} .
\label{turn1}
\eeq

If non-relativistic matter domination gives way directly to
positive cosmological-constant domination, then the scale-factor 
evolution can be written
\beq
\a(\tau) = \left(\frac{3\tc}{2\HL^2}\right)^{\!1/3}\,
e^{\HL\tau-2/3} \,,
\eeq
where in this context (i.e.~when there is no epoch of late-time 
spatial-curvature domination) the transition time is given by
$\tau_\Lambda =(2/3)\HL^{-1}$.  Whether or not there is a period of 
spatial-curvature domination between non-relativistic matter 
domination and cosmological-constant domination depends on whether 
or not this $\tau_\Lambda$ is greater than $\tc$; thus there is no 
epoch of late-time curvature domination when $\tc>(2/3)\HL^{-1}$.  
Proceeding, we write the above solution in terms of conformal time,
\beq
\a(\eta) = \HL^{-1}\bigg[
\left(\textstyle\!
\frac{1}{18}\HL\tc\right)^{\!-1/3}\!-\,\eta\bigg]^{-1}, 
\qquad 
\eta = \left(\textstyle\frac{3}{2}\HL\tc\right)^{\!-1/3}
\left(3-e^{-\HL\tau+2/3}\right) \,.
\eeq

When the cosmological constant is negative, we run into the same
problem as before.  Therefore, for this case we include both 
non-relativistic matter and cosmological constant in the Einstein 
field equation.  Since we are ultimately only interested in periods 
after non-relativistic matter domination in cosmologies with
$\tau_{\rm eq}\ll\HL^{-1}$, we can simply take the solution  
\beq
\a(\tau) = \left(\frac{3\tc}{2\HL^2}\right)^{\!1/3}
\sin^{2/3}\!\left(\frac{3}{2}\HL\tau\right) ,
\label{mtonl}
\eeq
which matches onto the late-time non-relativistic matter 
domination solution in the small-time limit.  The scale-factor 
solution in terms of conformal time can be approximated by combining 
(\ref{mtonl}) with the inversion of
\beq
\eta = -\left(\frac{2}{3\HL\tc}\right)^{\!1/3}\left\{
\frac{2\pi^{3/2}}{\Gamma(-\frac{1}{3})\,\Gamma(\frac{5}{6})}
+\frac{2}{3}\cos\!\left(\frac{3}{2}\HL\tau\right) F\!\left[
\frac{1}{2},\,\frac{5}{6},\,\frac{3}{2},\,
\cos^2\!\left(\frac{3}{2}\HL\tau\right)
\right] \right\},
\eeq
where $F$ is the hypergeometric function, and we have again used 
$\tau_{\rm eq}\ll\HL^{-1}$ to drop subleading terms.  As in the case 
where there is a period of spatial-curvature domination, negative 
cosmological-constant domination gives way to turnaround followed by 
the time reversal of the previous scale-factor evolution.  In this 
case the turnaround time is given by
\beq
\tau_{\rm turn} = \frac{\pi}{3\HL}\,, \qquad
\eta_{\rm turn} = -\frac{2\pi^{3/2}}{\Gamma(-\frac{1}{3})\,
\Gamma(\frac{5}{6})}\left(\frac{2}{3\HL\tc}\right)^{\!1/3}\,.
\label{turn2}
\eeq

\subsection{Growth factor evolution}

In the linear approximation, a spherical top-hat overdensity in 
non-relativistic matter obeys the equation of motion
(see for example \cite{Dodelson}) 
\beq
\ddot{\sigma}+2H\dot{\sigma}-\frac{9}{4}\frac{\tc}{\a^3}\sigma=0 \,,
\label{se}
\eeq
where $\sigma\equiv\delta\rho_{\rm m}/\rho_{\rm m}$ and we have 
used that the non-relativistic matter density can be written
$\rho_{\rm m} = (3/2)\tc/\a^3$.  The primordial density 
perturbations in bubbles like ours are nearly scale invariant 
and at least approximately Gaussian; we are interested in the 
root-mean-square (rms) density contrast averaged over a comoving 
scale enclosing a mass $M$ of non-relativistic matter.  It is 
customary to separate this quantity into two parts, writing
\beq 
\sigma_{\rm rms}(M,\tau) = \sigma_{\rm ref}(M)\,D(\tau) \,,     
\eeq
where $\sigma_{\rm ref}$ is the rms density contrast evaluated
at some reference time $\tau_{\rm ref}$, which we take to be 
deep into the period of non-relativistic matter domination, and 
$D$ is the growth factor, describing the evolution of $\sigma$ 
after that.  We fix $\sigma_{\rm ref}$ to match observation
\cite{WMAP}.

It is customary to use the linearized equations of motion to 
determine the growth factor $D$ even after perturbations have 
grown so large as to make the linear approximation inappropriate, 
and to account for this convention separately (see the main text).  
When the bubble can be approximated as containing only 
non-relativistic matter, spatial curvature, and cosmological 
constant, (\ref{se}) admits the solution \cite{Dodelson}
\beq
D(\a) = \frac{5}{2}\left(\frac{\tc}{\tau_{\rm ref}}\right)^{\!2/3}
\left(\frac{\gamma^3}{\HL^3\a^3}+\frac{1}{\HL^2\a^2}\pm 1\right)^{\!1/2}
\left[ C + \int_0^{\HL\a} dx 
\left(\frac{\gamma^3}{x}+1 \pm x^2\right)^{\!-3/2} \right] ,
\label{Dgen}
\eeq
where $\gamma\equiv (\frac{3}{2}\HL\tc)^{1/3}$, $\pm$ 
corresponds to the sign of the cosmological constant, and $C$ is an 
integration constant.  The constant $C$ is unimportant at times 
$\tau\gg\tau_{\rm ref}$, and can therefore be set to zero, except 
after scale-factor turnaround in bubbles with negative cosmological 
constant.  In that case $C$ is determined by demanding that $dD/d\tau$ 
is continuous at turnaround (the form of (\ref{Dgen}) guarantees that 
it is continuous at turnaround).  Formally, this gives
\bea
C &=& \lim_{\a\to\a_{\rm turn}} \Bigg\{ \!
\left(\frac{3}{4}\frac{\gamma^3}{\HL\a}+\frac{1}{2}\right)^{\!-1}\!
\left(\frac{\gamma^3}{\HL^3\a^3}+\frac{1}{\HL^2\a^2}- 1\right)^{\!-1/2}
\nn\\
& & \qquad\quad\,\, -\, 2\int_0^{\HL\a} dx 
\left(\frac{\gamma^3}{x}+1 - x^2\right)^{\!-3/2} \Bigg\} ,
\label{Cgen}
\eea
where $\a_{\rm turn}$ is the value of the scale factor at turnaround.
The divergence in the first term is canceled by a divergence in the
integral, when the expressions are regulated.  

Unfortunately, we cannot obtain a closed-form expression for the 
integral in (\ref{Dgen}).  Nevertheless, a bit of trial and error 
allows us to find a set of results that reasonably approximate $D$. 
To begin, we focus on bubbles with positive cosmological constant, 
and divide the integration over $x$ into two parts:  one for which 
the sum $\gamma^3/x+1$ dominates the integrand (corresponding to the
effects of non-relativistic matter and spatial curvature dominating 
the Hubble rate), the other for which the sum $1+x^2$ dominates the
integrand (corresponding to the effects of spatial curvature and 
cosmological constant dominating the Hubble rate).  These two
parts are matched at $x=\gamma$, which corresponds to when the 
energy density in non-relativistic matter is equal to the energy 
density in cosmological constant.  Accordingly, for 
$\HL\a< (\frac{3}{2}\HL\tc)^{1/3}$, in which case $x<\gamma$ over
the entire range of integration, we write 
\bea
D(\a) &=& \frac{5}{2}\left(\frac{\tc}{\tau_{\rm ref}}\right)^{\!2/3}
\left(\frac{\gamma^3}{\HL^3\a^3}+\frac{1}{\HL^2\a^2}\right)^{\!1/2}
\int_0^{\HL\a} dx 
\left(\frac{\gamma^3}{x}+1\right)^{\!-3/2} \nn\\
&=& \frac{5}{2}\left(\frac{\tc}{\tau_{\rm ref}}\right)^{\!2/3}
\Bigg\{ 1 +\frac{3\gamma^3}{\HL\a}-\frac{3\gamma^3}{\HL\a}
\left(1+\frac{\gamma^3}{\HL\a}\right)^{\!1/2}
{\rm arcsinh}\!\left[\frac{(\HL\a)^{1/2}}{\gamma^{3/2}}\right]\!
\Bigg\} \,, \label{Dcurv1}
\eea
where the prefactor $H(\a)/\HL$ has been modified from (\ref{Dgen}),
in accordance with the above approximation, to improve accuracy and 
provide continuity in the full result.  Meanwhile, for 
$\HL\a\geq (\frac{3}{2}\HL\tc)^{1/3}$ we write  
\bea
D(\a) &=& \frac{5}{2}\left(\frac{\tc}{\tau_{\rm ref}}\right)^{\!2/3}
\left(\frac{1}{\HL^2\a^2}+1\right)^{\!1/2}\Bigg\{
\int_0^{\gamma} dx
\left(\frac{\gamma^3}{x}+1\right)^{\!\!-3/2}\!+\int_{\gamma}^{\HL\a} dx
\left(\gamma^2+1+x^2\right)^{\!-3/2} \Bigg\} \nn\\
&=& \frac{5}{2}\left(\frac{\tc}{\tau_{\rm ref}}\right)^{\!2/3}
\left(\frac{1}{\HL^2\a^2}+1\right)^{\!1/2}
\Bigg\{ \frac{3\gamma^3+\gamma}{(\gamma^2+1)^{1/2}}-3\gamma^3
{\rm arcsinh}\!\left(\gamma^{-1}\right) \nn\\
& & +\, \frac{\gamma^{1/2}}{\gamma^2+1}
\left[\frac{\HL\a}{\big(\gamma^3+\gamma+\gamma\HL^3\a^3\big)^{1/2}}
-\frac{\gamma^{1/2}}{\big(2\gamma^2+1\big)^{1/2}}\right]\! \Bigg\} \,,
\eea
where we have added the term $\gamma^2$ to the second integrand
because it significantly improves the accuracy of the approximation.

The case of negative cosmological constant is treated similarly. 
Indeed, we do not significantly increase the error if we
compute $D(\a)$ as if the cosmic fluid contained only non-relativistic 
matter and spatial curvature all the way up to scale-factor turnaround 
(of course we continue to use the scale factor solution that includes 
the effect of negative cosmological constant).  This gives (\ref{Dcurv1}), 
now with the understanding that the result applies only for 
$\tau\leq\tau_{\rm turn}$.  For later times, we incorporate the 
integration constant $C$ mentioned above.  However, due to the present 
approximation, $D(\a)$ is not continuous at $\a_{\rm turn}$ for any 
given value of $C$, so we set $C$ to the particular value
for which the solution is continuous.  This gives, for 
$\tau>\tau_{\rm turn}$,
\bea
D(\a) &=& \frac{5}{2}\left(\frac{\tc}{\tau_{\rm ref}}\right)^{\!2/3}
\left(\frac{\gamma^3}{\HL^3\a^3}+\frac{1}{\HL^2\a^2}\right)^{\!1/2}
\Bigg\{ 
\frac{2+6\gamma^3}{\left(1+\gamma^3\right)^{1/2}}
-\left(\frac{\gamma^3}{\HL^3\a^3}+\frac{1}{\HL^2\a^2}\right)^{\!-1/2} \nn\\
& & -\, 3\gamma^3\left(1+\frac{\gamma^3}{\HL\a}\right)^{\!-1/2}\!\!
-6\gamma^3{\rm arcsinh}\big(\gamma^{-3/2}\big)
+3\gamma^3{\rm arcsinh}\!\left[\frac{(\HL\a)^{1/2}}{\gamma^{3/2}}\right]\!
\Bigg\} . \,\,\,\,
\eea

The case where non-relativistic matter domination gives way 
directly to cosmological-constant domination is included in the 
analysis above, as the limit of large $\gamma$.  For sufficiently 
large $\gamma$, however, it is worthwhile to consider a different 
approximation, which improves the accuracy, especially with respect
to the asymptotic value of $D$.  Since for large $\gamma$ spatial
curvature never contributes significantly to the post-inflationary 
energy density, to approximate this case we simply ignore the 
corresponding terms in (\ref{Dgen}).  This gives
\bea
D(\a) &=& \frac{5}{2}\left(\frac{\tc}{\tau_{\rm ref}}\right)^{\!2/3}
\left(\frac{\gamma^3}{\HL^3\a^3}\pm 1\right)^{\!1/2}
\int_0^{\HL\a} dx \left(\frac{\gamma^3}{x} \pm x^2\right)^{\!-3/2} \nn\\
&=& \left(\frac{\tc}{\tau_{\rm ref}}\right)^{\!2/3} 
\frac{\HL\a}{\gamma^3}
\left(1\pm\frac{\HL^3\a^3}{\gamma^3}\right)^{\!-1/2}
\Bigg\{ \frac{8}{3}\,F\!\left(-\frac{1}{2},\,\frac{5}{6},\,\frac{11}{6},\,
\mp\frac{\HL^3\a^3}{\gamma^3}\right) \quad\nn\\
& & -\left(\frac{5}{3}\pm\frac{2\HL^3\a^3}{3\gamma^3}\right)
F\!\left(\frac{1}{2},\,\frac{5}{6},\,\frac{11}{6},\,
\mp\frac{\HL^3\a^3}{\gamma^3}\right)\!
\Bigg\}\, ,
\label{Dnocurv}
\eea
where again $F$ denotes the hypergeometric function, and the upper 
(lower) entry of $\pm$ and $\mp$ corresponds to positive (negative) 
cosmological constant.  In the case $\Lambda<0$, the solution above
corresponds to $D$ before scale-factor turnaround.  The solution
after turnaround is obtained by incorporating an integration constant, 
as described below (\ref{Dgen}), except in the present approximation we 
drop the terms associated with spatial curvature.  This gives  
\bea
D(\a) &=& \frac{5}{2}\left(\frac{\tc}{\tau_{\rm ref}}\right)^{\!2/3}
\frac{\HL\a}{\gamma^3}\left(1-\frac{\HL^3\a^3}{\gamma^3}\right)^{\!-1/2}
\Bigg\{ \frac{4\sqrt{\pi}\,\Gamma(\frac{5}{6})}
{3\,\Gamma(\frac{1}{3})}\frac{\gamma}{\HL\a}
\left(\frac{\gamma^3}{\HL^3\a^3}+\frac{\HL^3\a^3}{\gamma^3}-2\right)^{\!1/2}
\nn\\
& & +\,
\frac{8}{3}\,F\!\left(-\frac{1}{2},\,\frac{5}{6},\,\frac{11}{6},\,
\frac{\HL^3\a^3}{\gamma^3}\right) 
-\left(\frac{5}{3}-\frac{2\HL^3\a^3}{3\gamma^3}\right)
F\!\left(\frac{1}{2},\,\frac{5}{6},\,\frac{11}{6},\,
\frac{\HL^3\a^3}{\gamma^3}\right)\!\Bigg\}\, .
\eea
Although the scale-factor solution of the previous section ignores the 
effect of spatial curvature for what corresponds to $\gamma\geq 1$,
in the present setting it increases the accuracy of the approximation 
to use (\ref{Dnocurv}) only for $\gamma\geq 1.9$ when $\Lambda > 0$. 

\begin{figure}[t!]
\begin{center}
\begin{tabular}{ccc}
\includegraphics[width=0.44\textwidth]{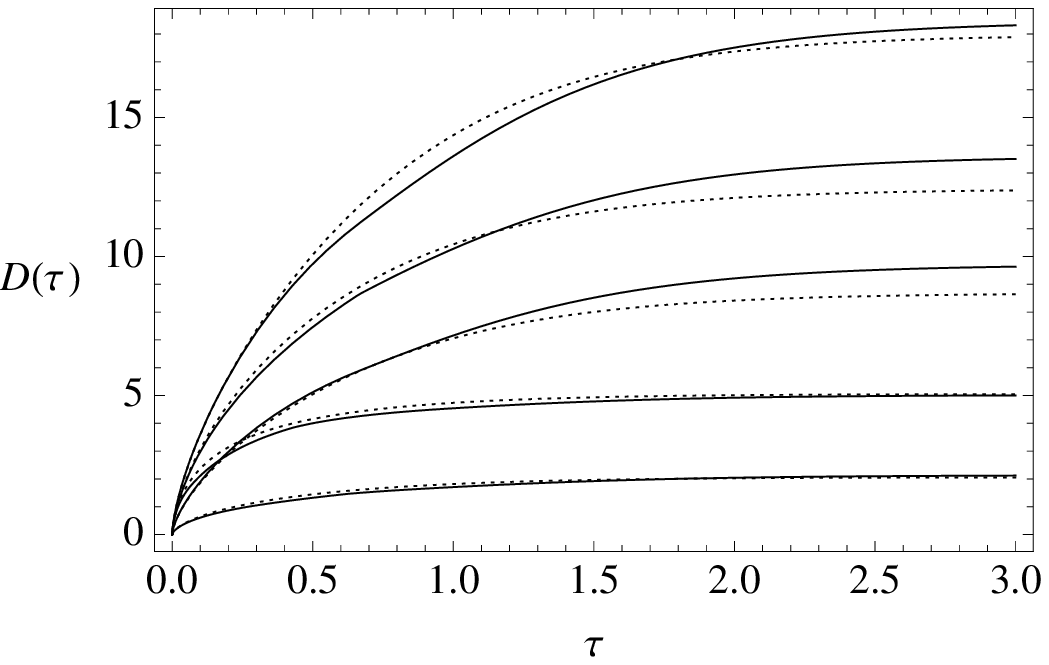} & &
\includegraphics[width=0.44\textwidth]{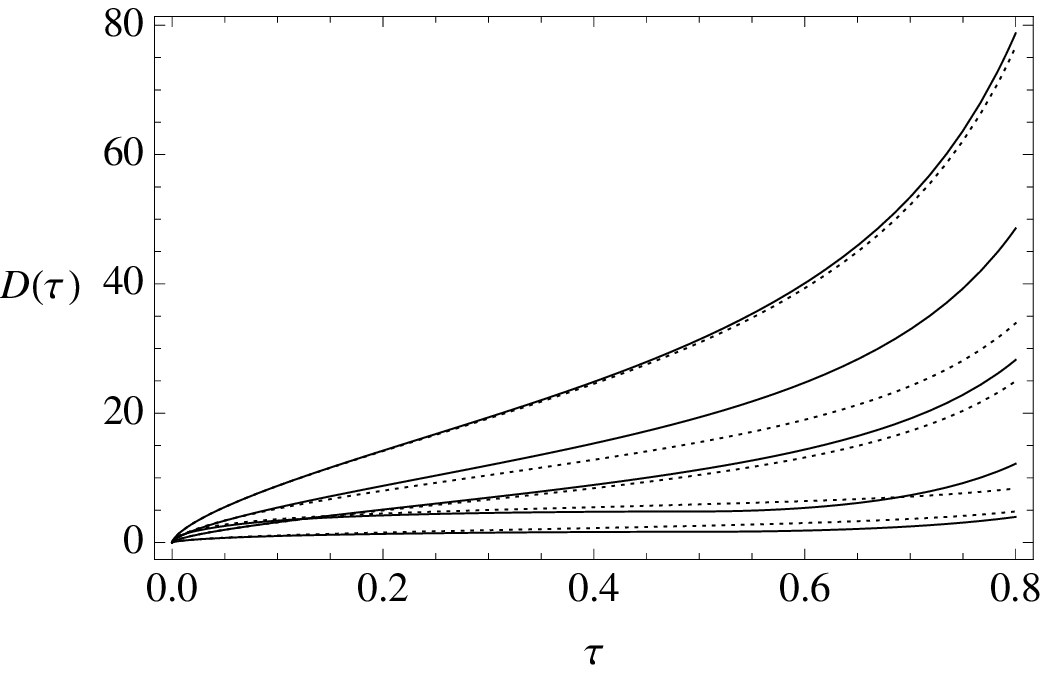}
\end{tabular}
\caption{\label{fig:D} Comparison of two methods to compute the 
growth factor $D(\tau)$, (1) using the approximations developed
in this appendix (solid curves) and (2) numerically integrating
the linear evolution equation (dotted curves), for positive (left 
panel, where time is given in units of $\HL^{-1}$) and negative 
(right panel, where time is given in units of $\tau_{\rm turn}$) 
cosmological constant, and several values of $\gamma^3$: 0.1 
(second from bottom), 0.5 (bottom), 2.0 (middle), 10
(second from top), and 100 (top).  Some pairs of curves been r
escaled to increase clarity.}
\end{center}
\end{figure}

It is not difficult to numerically check the accuracy of the above
approximations against (\ref{Dgen}), since aside from a common 
prefactor, they can all be expressed as functions of $x=\HL\a$, 
involving only one additional parameter $\gamma$.  The 
approximations are poorest for $\gamma\sim {\cal O}(1)$; several
representative curves from this region of the parameter space 
are displayed in Figure \ref{fig:D}.  Since some error is introduced 
by the approximations used to determine the scale-factor $\a$, in 
Figure \ref{fig:D} we compare two methods to determine $D(\tau)$ (as 
opposed to $D(\a)$).  One method combines (\ref{Dgen}) with the 
numerically-integrated Einstein field equation, 
\beq
\HL\tau = \int_0^{\HL\a} dx 
\left(\frac{\gamma}{x}+1+x^2\right)^{\!1/2} ,
\label{agen}
\eeq
inverted to give $\a(\tau)$.  The other is the method outlined 
above, using the approximations for $\a(\tau)$ described in the 
previous subsection.  Although the fits are not perfect, they are 
sufficient given the other uncertainties in our analysis, and 
they rapidly improve as $\gamma$ becomes very large or very small.  
(Note that because $\gamma \propto e^{N}$, the approximations are 
poorest only very near special values of $N$, and the curves of 
Figure \ref{fig:D} survey near those special values.) 

Throughout the main text we use only the closed-form approximations 
for the growth function $D(\a)$, in conjunction with the closed-form 
scale-factor solutions $\a(\tau)$ presented in the previous 
subsection, as opposed to the more computationally intensive 
numerical evaluations to which they are compared in Figure \ref{fig:D}.

\end{document}